\documentclass[preprint,aps,prc,showpacs,nofootinbib]{revtex4}
\usepackage{epsfig}
\usepackage{graphicx,amsmath}
\newcommand\ba{\begin{eqnarray}}
\newcommand\ea{\end{eqnarray}}

\newcommand{\be}{\begin{equation}}
\newcommand{\ee}{\end{equation}}

\begin{document}
\title{Testing axial and electromagnetic nucleon form factors in  time-like regions in the 
processes  $\bar p + n\to \pi^- +\ell^-+\ell^+$ and $\bar p+ p\to \pi^0 +\ell^- + \ell^+$, $\ell=e$, $\mu$ }

\author{C. Adamu\v s\v c\'in }
\altaffiliation{\it Department of Theoretical Physics, IOP, Slovak Academy of Sciences, Bratislava, Slovakia}
\affiliation{ DAPNIA/SPhN, CEA/Saclay, 91191 Gif-sur-Yvette Cedex, France}

\author{E. A. Kuraev}
\altaffiliation{\it JINR-BLTP, 141980 Dubna, Moscow region, Russian
Federation}
\affiliation{ DAPNIA/SPhN, CEA/Saclay, 91191 Gif-sur-Yvette Cedex, France}

\author{E. Tomasi-Gustafsson}
\email{etomasi@cea.fr}
\affiliation{ DAPNIA/SPhN, CEA/Saclay, 91191 Gif-sur-Yvette Cedex, France}

\author{F. E. Maas}
\affiliation{CNRS/INP3, IPN Orsay , 91400 Orsay Cedex, France}

\date{\today}
%\pacs{25.30.Bf, 13.40.-f, 13.40.Gp}

\begin{abstract}
In frame of a phenomenological approach based on Compton-like Feynman amplitudes, 
we study the annihilation channel in antiproton nucleon collisions with production  of a single 
charged or neutral pion and a lepton-antilepton pair. These reactions allow to access nucleon and 
axial electromagnetic form factors in time-like region and offer a unique possibility to study the  
kinematical region below two nucleon threshold. The differential cross section in an experimental set-up 
where the pion is fully detected is given with explicit dependence on the relevant nucleon form factors. 
The possibility to measure heavy charged pion in the annihilation channel is also discussed.
\end{abstract}

\maketitle
%%%%%%%%%%%%%%%%%%%%%%%%%%%%%%%%%%%%%%%%%%%%%%%%%%%%%%%%%%%%%%%%%
\section{Introduction}
%%%%%%%%%%%%%%%%%%%%%%%%%%%%%%%%%%%%%%%%%%%%%%%%%%%%%%%%%%%%%%%%%
\label{intro}
In this paper we study the annihilation reaction 
$\bar{p} + n\to \pi^- +\ell^-+\ell^+$ and $\bar p+ p\to \pi^0 +\ell^- + \ell^+$, $\ell=e$, $\mu$
which is the crossed process of pion electroproduction on a nucleon $N$: $e^- + N \to e^- +  N  + \pi$.
It contains the same information on the nucleon from factors in a different kinematical region.
This process is also related to the pion scattering process $\pi + N \to N + \ell^- + \ell^+$ which was first studied 
in \cite{Re65}. In this work it was already pointed out that the $\bar{p} + p$ annihilation process with pion production, under study here, 
renders possible the determination of the nucleon electromagnetic form factors in the 
unphysical region, which is otherwise unreachable in the annihilation reaction 
$\bar{p} + p \to e^+ + e^-$. In \cite{Du95}  a general expression 
for the cross section was derived and numerical estimations were given in the kinematical region
near threshold. In this paper we extend the formalism in two directions. We take into account
a larger set of diagrams which can contribute and give special emphasis to the possibility 
to access the axial nucleon form factors in the time-like region.
For the annihilation process $\bar p  N\to \ell^+ + \ell^-+\pi^{0,-}$, with $\ell=e,$ or $\mu$, 
we will consider two reactions: 
\be
\bar{p} (p_1)+p(p_2) \to \pi^0(q_{\pi})+\ell^+(p_+)+\ell^-(p_-), 
\label{eq:eqp}
\ee
and 
\be
\bar{p} (p_1)+n(p_2) \to \pi^-(q_{\pi})+\ell^+(p_+)+\ell^-(p_-), 
\label{eq:eqn}
\ee
where the notations of the particle four momenta are indicated in brackets. The interest of these processes lies in the possibility to access 
nucleon  and axial form factors (FFs) in the time-like region. The  expected values of the total cross sections at energies of a few GeV 
drop with the lepton pair invariant mass $q^2$, but are of the order of several nb and up to mb below threshold.  
Therefore these reactions are especially interesting as they will be measurable in next future at hadron colliders (or the crossed reactions at lepton colliders). 
The present work aims to evaluate the differential cross section for experimental conditions achievable at the future FAIR facility \cite{FAIR}.

%%%%%%%%%%%%%%%%%%%%%%%%%%%%%%%%%%%%%%%%%%%%%%%%%%%%%%
\subsection{Electromagnetic (Vector) Form Factor}
The structure of proton and nucleon is of great interest since it explores 
the quantum field theory of strong interaction, quantum chromo dynamiscs (QCD),
in a region where the interaction between the constituents of quarks and gluons 
can not be treated as a perturbation. Its properties have been 
explored in many observables. For example, the electromagnetic form factors of 
proton and nucleon are measured in elastic electron-nucleon scattering with a space-like 
four momentum transfer  $q^2 < 0 $
via the process $e + p \to e + p$ \cite{paper:Hofstadter}. For a momentum transfer of $-1$ GeV$^2$ $< q^2 <0$ GeV$^2$,  
the electromagnetic form factors are well known for both neutron and proton \cite{paper:friedrichwalcher:formfactors}.  

The interest in the form factors of the nucleon for the region of $ q^2< -1$ GeV$^2$
has recently been renewed by a measurement 
of the ratio of electric form factor over the magnetic from factor $\mu_{p}$$G_{E}$/ $G_{M}$ at TJNAF 
in a $q^2$-range up to - 5.8 GeV$^2$. The method of polarization transfer \cite{Re68} was employed for the first time at high values of $q^{2}$ using 
a longitudinally polarized electron beam and measuring the polarization of the recoil proton \cite{Jo00}.
The experiments gave the surprising new result that the ratio  $\mu_{p}$$G_{E}$/ $G_{M}$ decreases 
from unity substantially for high $q^2$-values exhibiting eventually even a zero-crossing
around $q^2 \approx - 8$ GeV$^2$. This new result is in contradiction with all data 
obtained from a Rosenbluth separation fit of unpolarized measurements
\cite{paper:rosenbluthformfactors} giving a ratio of $\mu_{p}G_{E}/G_{M}\approx 1$. 
The present understanding of this discrepancy is that the analysis of the unpolarized data lacks 
important corrections stemming from neglected higher order radiative corrections \cite{paper:radiativecorrections}
or electromagnetic processes like two-photon-exchange \cite{paper:twophotoncorrections}.

The situation of the experimental determination of the electromagnetic form factors in the time-like domain 
(i.e. $q^{2} > 0$) is quite different. Existing data have explored the two basic processes:
 $e^{-} + e^{+} \to p + \bar{p}$ and  $p +\bar{p} \to e^{-} + e^{+}$. 
The precision of the cross section data in the time-like region is much lower and a 
determination of the individual electric and magnetic form factors has not been done with sufficient precision.\\
For both processes there is a  threshold in $q^2$  for producing the two nucleons at rest or 
annihilation of the two nucleons at rest which amounts to $q^2 > (m_{p} + m_{\bar{p}})^2 = 4 m_{p}^2$.
In the region of $0 < q^2 < 4 m_{p}^2$ (sometimes called the  ``unphysical region'') 
there are no data available. 
On the other hand this $q^2$-interval  is of great interest since the intermediate virtual photon as well 
as the $\bar{p}$$p$-pair can couple to vector meson- and $\bar{p}$$p$-resonances respectively
and thereby enhance the form factors substantially \cite{paper:formfactors:dispersionrelation,Ba06}.
This mechanism has been used to explain the large cross section in the time-like region 
above the threshold $q^2 >  4 m_{p}^2$.
Another possibility to explain this large cross section above threshold has 
been used recently by showing, that the large cross section in the annihilation process
$\bar{p} + p \to e^+ + e^- \label{eq:elastictimelike}$ can be related to the $\bar{p}$$p$ scattering length
\cite{paper:meissner_hammer:ppbarscatteringlength}. 

The work presented here
is focused on a possibility of extracting the electromagnetic form factors of the nucleon
in the ``unphysical region'' and above using the process $\bar{p} + p \to \pi + e^+ + e^- $,
as it has been proposed earlier \cite{Re65, Du95}. 
The measurement of the cross section of this reaction will be accessible at the future FAIR facility at GSI.

%%%%%%%%%%%%%%%%%%%%%%%%%%%%%%%%%%%%%%%%%%%%%%%%%%%
\subsection{Axial Form Factor}
In addition, we explore here also the new idea of accessing the axial vector current of the nucleon.
The axial form factors in the space-like region are measured in nuclear $\beta$-decay,
in neutrino scattering, in muon capture and in pion-electroproduction
($e + p \to e + n + \pi^+ $). A review on the present status of the 
axial structure of the nucleon in the space-like region is given in Ref.~\cite{paper:ga-review}. 
The first methods represent a rather direct measurement in the sense 
that the axial coupling of the weak charged currents is used to measure the axial form factor.
The extraction of the axial form factor from pion-electroproduction in the space like region is 
possible due to the application of a chiral Ward identity referred to as the 
Adler-Gilman relation \cite{paper:SchererKochAxialFF}. A recent review 
on the theoretical development can be found in Ref.~\cite{paper:FuchsSchererAxialFF}.
Corrections to order $\mathcal{O}(p^4)$ have been calculated in the framework 
of chiral perturbation theory in Ref.~\cite{paper:BernardKaiserMeissneraxialff}.
There are no data available on the axial form factor in the time-like region.
A direct measurement would be possible by studying the weak neutral or charged current
in the annihilation of $\bar{p}$$p$. Such a measurement is not accessible 
with present day experimental techniques. On the other hand an application of the 
Adler-Gilman relation to the matrix element of the crossed channel of pion-electroproduction,
namely $\bar{p} + n \to \pi^{-} + e^+ + e^-$, renders the possibility of accessing
the axial current also in the time-like region around and below threshold.
Despite the theoretical uncertainties, namely on one hand the applicability range of the 
Adler-Gilman relation which is strict only at the $\pi$-threshold and on the other 
hand the treatment of the off-shell nucleon between the $\pi$-vertex and the 
virtual photon vertex, the measurement of the cross section in  $\bar{p} + n \to \pi^{-} + e^+ + e^-$
at low energies would render the first estimation of the axial form factor in the 
time-like region. Also here, the measurement of the cross section of this reaction will
be accessible at the future FAIR facility at GSI.

\subsection{Approach}
%Recently, using an approach based on Chiral perturbation theory (ChPT) \cite{Fu03} the cross section of pion production was written in terms
%of Lagrangian parameters up to the order $p^3$. This analysis includes about ten unknown constants. Relations among them can be built and 
%connected to data from definite experiments. The formalism of ChPT is appropriate when the characteristic momenta of the problem do not 
%exceed ${4\pi f_{\pi}}$, $f_{\pi}=94 \mbox{~MeV}$. In the present case the incident antiproton momentum can be as high as 15 GeV. 
%Therefore if one wants to apply ChPT formalism, terms with high order momenta (up to $p^6$), should be included, introducing a large 
%number of additional (unknown) parameters.
%
Our approach is based on Compton-type annihilation Feynman amplitudes and aims to establish the matrix element of the processes 
(\ref{eq:eqp}) and (\ref{eq:eqn}). The main uncertainty in our description in terms of Green functions of mesons and nucleons 
(and their expected partners) is related to the model dependent description of hadron FFs and to the modeling of excited hadronic states.

The paper is organized as follows: the formalism is developed in Section II and some of the kinematical constraints for the considered 
reactions are discussed in Section III. Section IV contains a discussion on nucleon FFs and of our choices of parametrization, for
electromagnetic as well as for axial FFs. In Section V the numerical results for the differential cross section of the considered
processes will be presented. In Conclusions we discuss the results and summarize the perspectives opened by the experimental 
study of these reactions, including possible manifestation of heavy (radial) excited $\pi$ states.

%%%%%%%%%%%%%%%%%%%%%%%%%%%%%%%%%%%%%%%
\section{Formalism}
%%%%%%%%%%%%%%%%%%%%%%%%%%%%%%%%%%%%%%%

\begin{figure}
\begin{center}
\includegraphics[width=12cm]{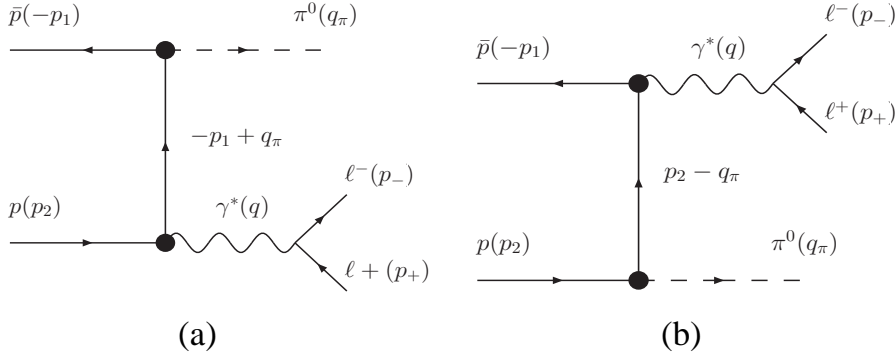}
\caption{\label{Fig:pp}Feynman diagrams for the reaction $ \bar{p} +p \to \pi^0+\ell^++\ell^-$.}
\end{center}
\end{figure}

\begin{figure}
\begin{center}
\includegraphics[width=12cm]{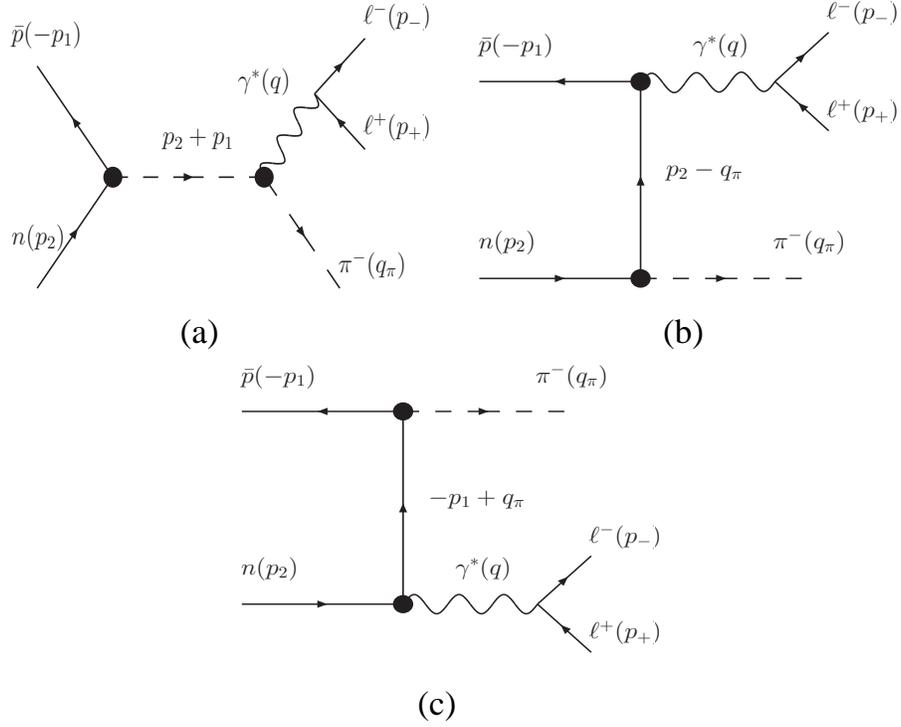}
\caption{\label{Fig:np}Feynman diagrams for the reaction $ \bar{p} +n \to \pi^-+\ell^++\ell^- $.}
\end{center}
\end{figure}

Let us consider the reactions (\ref{eq:eqp}) and (\ref{eq:eqn}) and calculate the corresponding matrix elements in 
the framework of a phenomenological 
approach based on Compton-like Feynman amplitudes. The Feynman diagrams for the reaction 
(\ref{eq:eqp}), are shown in  Fig. \ref{Fig:pp}a and \ref{Fig:pp}b, for pair emission from the proton and from the antiproton, respectively. 
For the reaction (\ref{eq:eqn}), the corresponding Feynman diagrams are shown in Figs. \ref{Fig:np}a, \ref{Fig:np}b and \ref{Fig:np}c for 
pair emission from the charged pion, from the antiproton and from the neutron, respectively.  

As previously discussed in Ref. \cite{Du95}, no free nucleons are involved in the electromagnetic vertexes of pions and 
nucleons, as one of the hadrons is virtual, and rigorously speaking, the form factors involved should be modified taking 
into account off mass shell effects. However we will use the expression for electromagnetic current involving on mass 
shell hadrons. A discussion of the errors and the consequences of such approximation is given below.

The vertexes $\gamma^* \bar N \bar N^*\to \gamma^* \bar N\bar N$ and 
$\gamma^* \pi \pi\to \gamma^* \pi \pi^*$ contain the information on the electromagnetic form factors (FFs) 
of proton, neutron and pion (FFs of antiproton differ by sign from proton FF, due to charge symmetry requirements). 
Nucleon FFs can be expressed in terms of Dirac and Pauli FFs $F_{1,2}^{p,n}(q^2)$, which enter in the expression 
of the electromagnetic current:
\be
<N(p')|\Gamma_\mu(q)^N|N(p)>=\bar u(p')\left [
F_1^N(q^2)\gamma_\mu+\frac{F_2^N(q^2)}{4M}(\hat{q}\gamma_\mu-\gamma_\mu\hat{q})
\right ] u(p),~N=n,p,
\ee
where $M$ is the nucleon mass and $q$ is the four-momentum of the virtual photon.
The nucleon FFs in the kinematical region of interest for the present work are largely unexplored. The assumptions 
and the parametrizations used for FFs in the numerical applications are detailed below.

%

%%%%%%%%%%%%%%%%%%%%%%%%%%%%%%%%%%%%%%%%%%%%%%%%%%%%%%%%%%%%%%%%%%%%%%%%%%%
%%%%%%%%%%%%%%%%%%%%%%%%%%%%%%%%%%%%%%%%%%%%%%%%%%%%%%%%%%%%%%%%%%%%%%%%%%%%%

%%%%%%%%%%%%%%%%%%%%%%%%%%%%%%%%%%%%%%%%%%%%%%%%%%%%%%%%%%%%%%%%%%%%%%%%%%
%%%%%%%%%%%%%%%%%%%%%%%%%%%%%%%%%%%%%%%%%%%%%%%%%%%%%%%%%%%%%%%%%%%%%%%%%%%%%
The pion electromagnetic FF, $F_{\pi}(q^2)$ is also introduced in the standard way. The corresponding current has the form:
\be
J_{\mu}^{\pi}=(q_1+q_2)_{\mu} F_{\pi}(q_{\pi}^2)
\ee
with $q_1$ and $q_2$ ingoing and outgoing charged pion momenta, and $q_{\pi}=q_1- q_2$.
Special attention must be devoted to the pion nucleon interaction, in the vertexes $\pi N\bar N$ that are parametrized as:
\be
\bar v(p_1)\gamma_5 u(p_2) g_{\pi NN}(s), \mbox{~and~}
\bar v(p_1-q)\gamma_5 u(p_2) g_{\pi NN}(m_{\pi}^2).
\ee
with $s=(p_1+p_2)^2$. The vertex of $\pi NN$ interaction is related to the general axial vector current matrix element:
\be
<N(p')|A_j^{\mu}(0)|N(p)>=\bar u(p')\left [G_A(q^2)\gamma_{\mu}+
\frac{q^{\mu}}{2M} G_P^2(q^2)+ i \displaystyle\frac{\sigma^{\mu\nu}}{2M} G_T(q^2)\right ] \gamma_5 \frac{\tau _j}{2} u(p),
\ee
where $q_{\mu}=p'_{\mu}-p_{\mu}$, $G_A(q^2)$ is the axial nucleon FF, $G_P(q^2)$ the induced pseudoscalar FF, 
and $G_T(q^2)$ the induced pseudotensor FF. In the chiral limit, the requirement of conservation of the axial current leads to the relation:
\be
4 M G_A(q^2) +q^2 G_P(q^2)=0,
\label{eq:ax1}
\ee
which shows that $G_P(q^2)$ has a pole at small $q^2$. Indeed, assuming that the axial current
interacts with the nucleon through the conversion to pion interaction, one obtains:
\be
G_P(q^2)= -\displaystyle\frac{4M f_{\pi} g_{\pi NN}(q^2)}{q^2}
\label{eq:ax2}
\ee
Comparing Eqs. (\ref{eq:ax1}) and (\ref{eq:ax2}) one obtains the Golberger-Treiman relation:
\be \displaystyle\frac{G_A}{f_{\pi}}= \displaystyle\frac{ g_{\pi NN}}{M}.
\label{eq:ax3}
\ee
We suggest a generalization of this relation in the form:
\be 
g(s)=g_{\pi\bar N N}(s)=\displaystyle\frac{MG_A(s)}{f_{\pi}},~G_A(0)=1.2673\pm 0.0035.
\ee
where
$g(s)$, $g(m_\pi^2)$ are the  pion-nucleon coupling constants
for pion off and on mass shell. This assumption can be justified by the fact that $f_{\pi}$ is weakly depending
on $q^2$ and it is in agreement with the ChPT expansion at small $q^2$ \cite{paper:FuchsSchererAxialFF}. Therefore measuring the 
$g_{\pi\bar N N}(s)$ coupling constant gives information on the axial and induced pseudoscalar FFs of the 
nucleon in the chiral limit (neglecting the pion mass).

The matrix element is expressed in terms of the hadronic $H$ and leptonic $J$ currents:
\be
M^i=\frac{4\pi\alpha}{q^2}H^i_\mu J^\mu(q),~H^i_\mu=\bar{v}(p_1)O^i_\mu u(p_2),~J^\mu(q)=\bar{v}(p_+)\gamma_\mu u(p_-), 
\ee
where the index $i=0,-$ refers to $\pi^0$ and $\pi^-$ respectively. The cross section for the case of 
unpolarized particles has a standard form (we imply that the nucleon target (proton or neutron) is 
at rest in the Laboratory frame):
\be
d\sigma^i=\frac{1}{16 PM}\sum|M^i|^2d\Gamma,~P^2=E^2-M^2,
\label{eq:sigma}
\ee
where $E$ ($P$) is the energy (the modulus of the momentum), and $d\Gamma$ is the phase space volume:
\be
d\Gamma=\frac{1}{(2\pi)^5}\frac{d^3p_+}{2\epsilon_+}\frac{d^3p_-}{2\epsilon_-}
\frac{d^3q_{\pi}}{2E_{\pi}}\delta^4(p_1+p_2-p_+-p_- -q_{\pi}).
\label{eq:gamma}
\ee
The phase space volume can be written as:
$$
d\Gamma=\frac{d^3q_{\pi}}{2E_{\pi}}d\Gamma_q 
\frac{d^4q}{(2\pi)^5} \delta^4(p_1+p_2-q-q_{\pi}),$$
with
$$ d\Gamma_q = \frac{d^3p_+}{2\epsilon_+}\frac{d^3p_-}{2\epsilon_-}
\delta^4(q -p_+ - p_-).$$
Considering an experimental set-up where the pion four-momentum is fully measured,  we can perform 
the integration on the phase space volume of the lepton pair:
\be
\int d\Gamma_q 
\sum J_\mu(q)J^*_{\nu}(q)=
-\frac{2\pi}{3}(q^2+2\mu^2)\beta\Theta(q^2-4\mu^2)
\left (g_{\mu \nu}-\frac{q_\mu q_\nu}{q^2}\right ),
\ee
where $\Theta$ is the usual step function, $\mu$ is the lepton mass and $\beta=\sqrt{1-(4\mu^2/q^2)}$ .

The cross section can be expressed in the form:
\be
d\sigma^i=\frac{\alpha^2}{6s\pi r}\frac{\beta(q^2+2\mu^2)}
{(q^2)^2}{\cal D}^i\frac{d^3q_\pi}{2\pi E_\pi},
\label{eq:dsig}
\ee
with
\be
s=(q_\pi+q)^2=2M(M+E), ~r=\sqrt{ 1-(4M^2/s)}
\label{eq:eqr}
\ee
 and 
\be
{\cal D}^i=\left (g_{\mu\nu}-\displaystyle\frac{q_\mu q_\nu}{q^2}\right )
\frac{1}{4}Tr(\hat{p}_1-M)O^i_\mu(\hat{p}_2+M)(O^i_\nu)^*,~ i=0,-.
\label{eq:eqcd}
\ee
Using Feynman rules we can write (see Figs. \ref{Fig:pp}, \ref{Fig:np}):
\begin{eqnarray}
O^-_\mu&=&\Gamma_\mu^p(q)\frac{\hat{p}_1-\hat{q}-M}{(p_1-q)^2-M^2}\gamma_5g(m_\pi^2)-\nonumber \\
&&
\gamma_5\frac{\hat{p}_2-\hat{q}+M}{(p_2-q)^2-M^2}
\Gamma_\mu^n(q) g(m_\pi^2)+\frac{(2q_\pi+q)_\mu}{s-m_\pi^2}g(s)F_{\pi}(q^2)\gamma_5,\label{eq:csnp}\\
O^0_\mu&=&\Gamma_\mu^p(q)\frac{\hat{p}_1-\hat{q}-M}{(p_1-q)^2-M^2}\gamma_5g(m_\pi^2)-\gamma_5
\frac{\hat{p}_2-\hat{q}+M}{(p_2-q)^2-M^2}
\Gamma_\mu^p(q) g(m_\pi^2);\label{eq:cspp}
\end{eqnarray}

Note that the hadronic current ${\cal J}_{\mu}^0=\bar v (p_1)O^0_{\mu}u(p_2)$ is conserved $J_{\mu}^0q^{\mu}=0$, 
but  $J_{\mu}^-= \bar v(p_1)O^-_{\mu}u(p_2)$ is not conserved:
\be 
q_{\mu} {\cal J}_{\mu}^-=\left [(-F_1^p(q^2)+F_1^n(q^2))g(m^2_{\pi})+g(s)F_{\pi}(q^2)\right ]
\bar v (p_1)\gamma_5 u(p_2)={\cal C} \bar v (p_1)\gamma_5 u(p_2).
\label{eq:cont}
\ee
Therefore, to provide gauge invariance, it is necessary to add to $O^-_{\mu}$ a contact term with the appropriate 
structure (\ref{eq:cont}). The explicit expressions for ${\cal D}^0 $ and ${\cal D}^- $ are given in the Appendix.

Selecting the coefficients which depend on pion energy in ${\cal D}_i$, Eq. (\ref{eq:eqcd}), an analytical integration on the pion energy can be performed. In the limit of small lepton pair invariant mass, $q^2$, after integration on pion energy, the 
differential cross section with respect to $q^2$ becomes:
\be 
(q^2)^2\left .\displaystyle\frac{d\sigma}{dq^2}\right |_{q^2\ll M^2}
\simeq \displaystyle\frac{\alpha^2 [g(s) -g(m_{\pi}^2)]^2}{24 \pi r }. 
\label{eq:eqy}
\ee
Eq. (\ref{eq:eqy}) contains one of the most important results of this work, as it shows that the measurement 
of the cross section at small $q^2$ allows to determine experimentally the off mass shell pion nucleon 
coupling constant.

Writing the differential cross section in the form:
\be
\displaystyle\frac{d\sigma}{dq^2}= \displaystyle\frac{(q^2+2\mu^2)\beta}{(q^2)^2}
\left [\displaystyle\frac{c}{q^2}+R(q^2)\right ], 
\ee
with $c$  and $R(0)$ finite functions of $s$. After integration of lepton invariant mass, we find:
\be
\sigma_{tot}=\int_{4\mu^2}^s\displaystyle\frac{d\sigma}{dq^2}dq^2= 
\displaystyle\frac{c(s)}{5\mu^2}+R(0,s)\left (\log \displaystyle\frac{s}{\mu^2}- 
\displaystyle\frac{5}{3}\right )+ \int_0^s\displaystyle\frac{dq^2}{q^2} [R(q^2,s)-R(0,s)].
\label{eq:eq23}
\ee
The first term in the right hand side of Eq. (\ref{eq:eq23}) is divergent for massless leptons, 
and induces a rise of the cross section (especially in the case of electron positron pair). 
However it is very hard to achieve experimentally such kinematics, $q^2\to 4 \mu^2$. The total cross section can be integrated within the experimental limits of detection of the particles. Such  (partial) total cross section will be calculated below.

%%%%%%%%%%%%%%%%%%%%%%%%%%%%%%%%%%%%%%%%%%

\section{Kinematics}
%%%%%%%%%%%%%%%%%%%%%%%%%%%%%%%%%%%%%%%%%%
In the Laboratory system, useful relations can be derived between the kinematical variables, 
which characterize the reaction. The allowed kinematical region, at a fixed incident total energy $s$
can be  illustrated as a  function of different useful variables.
 
One can find the following relation between $q^2$, the invariant mass of the lepton pair and the pion energy:
\be
q^2=(p_1+p_2-q_{\pi})^2=2M^2+m_{\pi}^2+2M(E-E_{\pi})-2p_1 q_{\pi}=
s+m_{\pi}^2-2E_{\pi}M  -2p_1 q_{\pi},
\label{eq:eq1}
\ee
with
\be
2p_1q_{\pi}=2E_{\pi}E- 2\sqrt{E_{\pi}^2-m_{\pi}^2} P\cos\theta_{\pi},
\label{eq:eq2}
\ee
where $\theta_{\pi}=\widehat{\vec p_1 \vec q_\pi}$ is the angle between the antiproton and the pion momenta 
(in the Laboratory frame). 

\begin{figure}
\begin{center}
\includegraphics[width=0.49\textwidth]{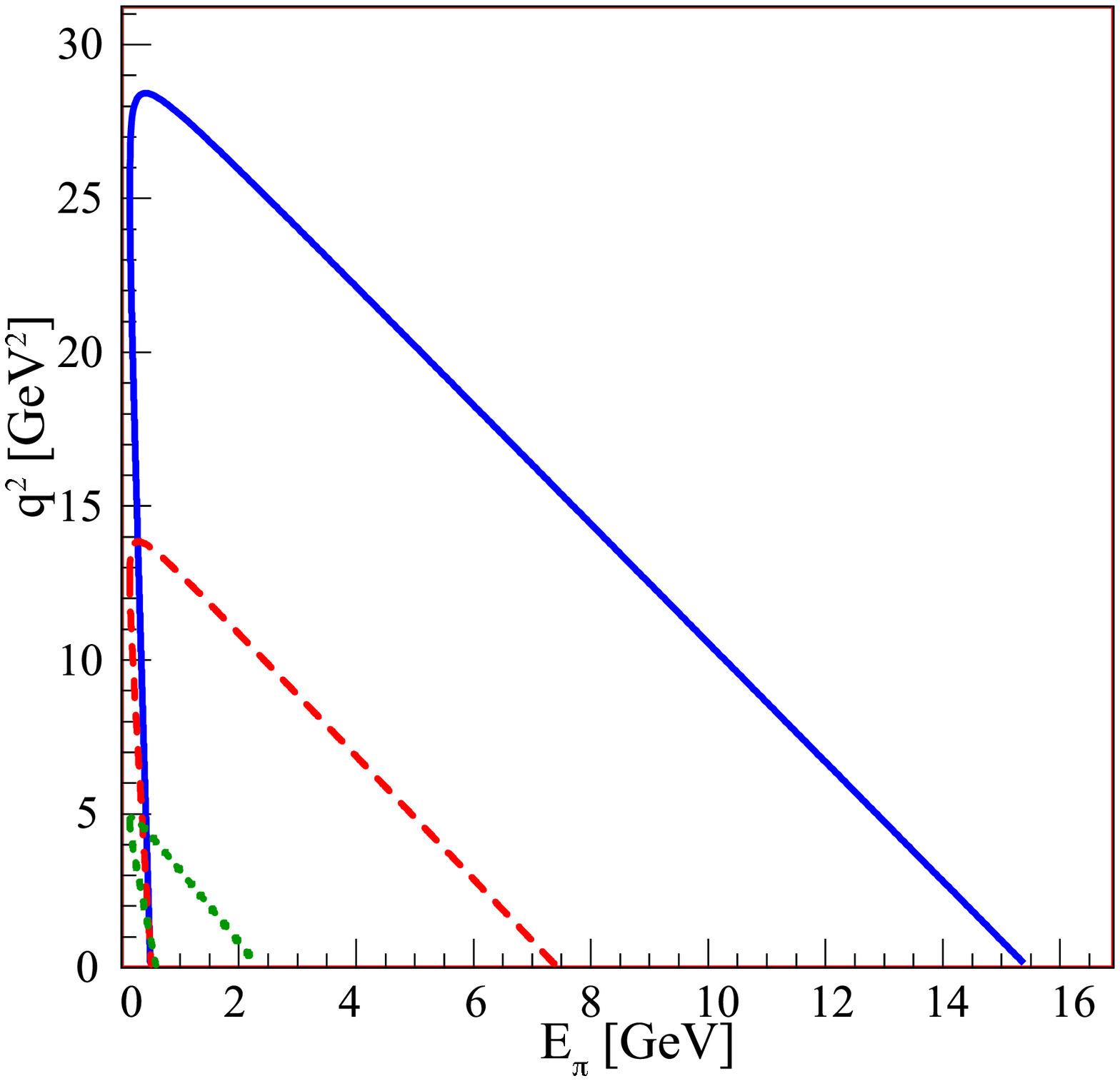}
\includegraphics[width=0.49\textwidth]{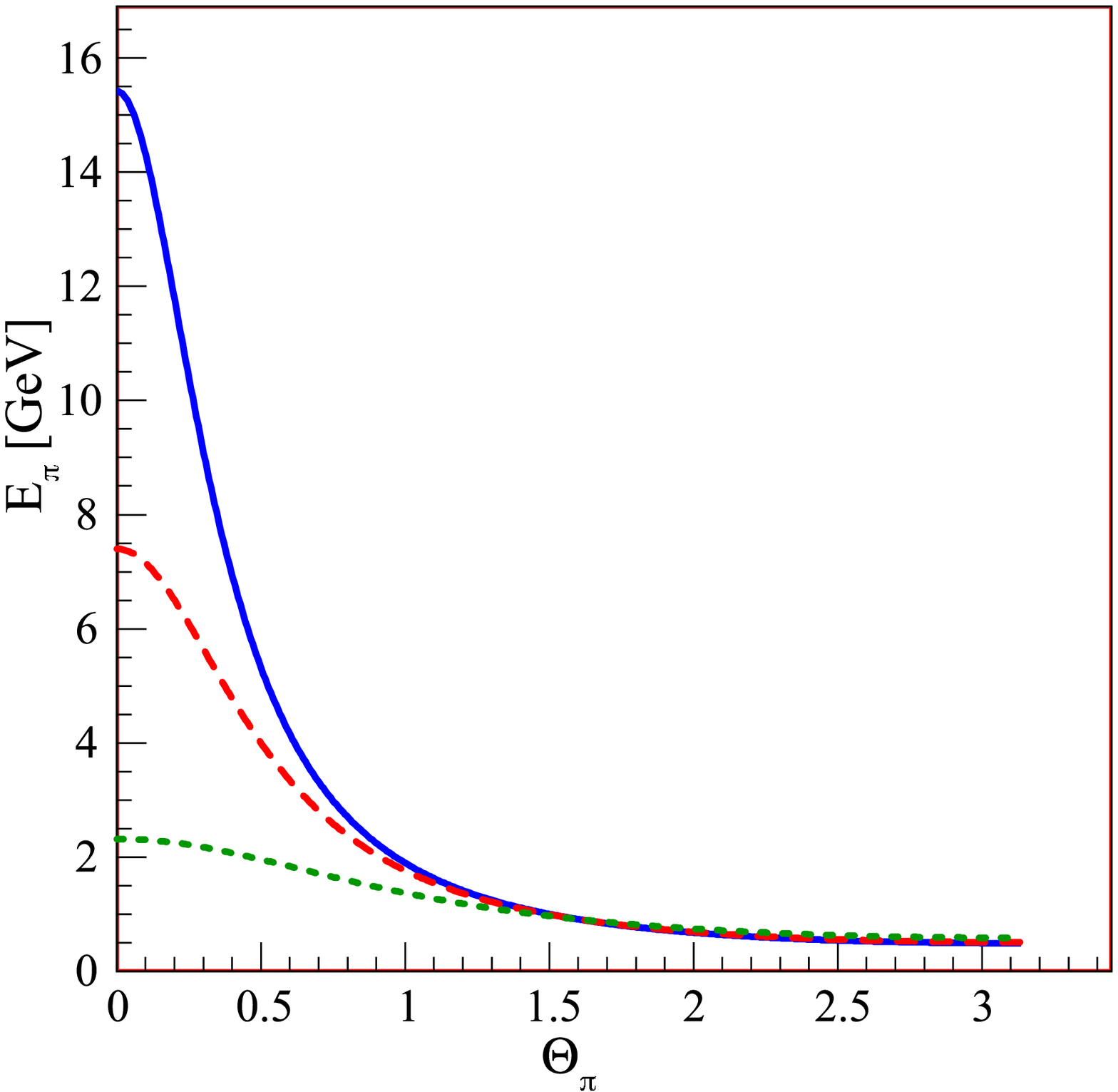}
\caption{\label{Fig:c2} (Color online) Left: The kinematical limit for $q^2$ is shown for 
$\cos\theta_{\pi} = -1$ and for $\cos\theta_{\pi} = 1$ in the Lab system as function 
of the pion energy for different values of the beam energy:  $E$=2 GeV$^2$ (dotted line), 
$E$=7 GeV$^2$ (dashed line), $E$=15 GeV$^2$ (solid line). 
The allowed kinematical region lies below the curves.\\
Right: Kinematical limit for the pion energy $E_{\pi}$ as a function of the pion
angle (Lab system), for $E$=2 GeV$^2$ (dotted line), $E$=7 GeV$^2$ (dashed line), $E$=15 GeV$^2$ (solid line), for the minimum value of $q^2\simeq m_\pi^2$.}
\end{center}
\end{figure}

The limit $-1\le \cos\theta_{\pi}\le 1$ translates into a maximal and a minimal value for the pion energy. 
The allowed kinematical region is shown in Fig. \ref{Fig:c2} left, for three values of the beam energy: $E =2$, 7, 15 GeV.
To this constraint, one should add the minimal thresholds $q^2\ge 4m_{\ell}^2$ and $ E_{\pi}\ge m_{\pi}$. 
For the minimal value of $q^2\simeq m_\pi^2$ , one can plot the dependence of the  pion energy on  
$\theta_{\pi}$ (Fig. \ref{Fig:c2}, right), for different values of the beam energy.
As the energy increases the kinematically allowed region becomes wider. At backward angles the 
maximum pion energy becomes larger at small $s$ values. For larger values of $q^2$, $E_{\pi}$ is smaller.

%%%%%%%%%%%%%%%%%%%%%%%%%%%%%%%%%%%%%%%%%%%%%%%%%%%%%%%%%%%%%%%%%%%%%%%%%%%
%%%%%%%%%%%%%%%%%%%%%%%%%%%%%%%%%%%%%%%%%%%%%%%%%%%%%%%%%%%%%%
For fixed values of the lepton pair invariant mass, the pion energy can take values in the region:
\be
\displaystyle\frac{E_{\pi}^{min}}{M}=
\displaystyle\frac{s-q^2}{s(1+r)}
\le \displaystyle\frac{E_{\pi}}{M}
\le \displaystyle\frac{s-q^2}{s(1-r)}
=\displaystyle\frac{E_{\pi}^{max}}{M}, 
\label{eq:eqm}
\ee
neglecting the pion mass.
%%%%%%%%%%%%%%%%%%%%%%%%%%%%%%%%%%%%%%%%%%%%%%%%%%%%%%%%%%%%%%%%%%%%%%%%%%%%%

%

%%%%%%%%%%%%%%%%%%%%%%%%%%%%%%%%%%%%%%%%%%%%%%%%%%%%%%%%%%%%%%%%%%%%%%%%%%%%%%%%%%%%%%%%%%%

The phase space volume of the produced pion can be written  (neglecting terms $\simeq m_{\pi}^2 /m_N^2$) in three (equivalent) forms:
\ba
\frac{d^3q_i}{2\pi E_\pi }&= &dq^2 \delta[q^2 -2 E_\pi (E+M-P\cos\theta_{\pi})]
E_\pi dE_\pi d\cos\theta_{\pi}= \nonumber\\
&=&E_\pi dE_\pi d\cos\theta_{\pi}  \hspace*{11.5 true cm} (27a )\nonumber\\
&=&M \frac{dq^2}{sr}dE_\pi  \hspace*{12. true cm} (27b)\nonumber\\
&=& \frac{q^2  M^2 dq^2d\cos\theta_{\pi} }{s^2(1-r\cos\theta_{\pi})^2} \hspace*{11 true cm} (27c)\nonumber 
\ea

%
%%%%%%%%%%%%%%%%%%%%%%%%%%%%%%%%%%%%%%%%%%%%%%%%%%%%%%%%%%%%%%%%%%%%%%%%%%%%%
\section{Axial and electromagnetic form factors}
%%%%%%%%%%%%%%%%%%%%%%%%%%%%%%%%%%%%%%%%%%%%%%%%%%%%%%%%%%%%%%%%%%%%%%%%%%%%%

Experimental measurements on  FFs are object of ongoing programs in several world facilities. Hadron FFs are measured 
in space-like (SL) region through elastic electron hadron scattering, and in time-like (TL) region through annihilation reactions. 
It has been only recently possible to use polarization techniques. The availability of high intensity, high energy polarized beams
 allows to extend these measurements to large $q^2$ regions and 
to achieve very high precision.

The theoretical effort for a complete description of the nucleon structure should be extended to a unified picture which 
applies to the full kinematical region (SL and TL) \cite{Etg01}. Few phenomenological models, developed for the 
SL region can be successfully extended to the TL region \cite{Etg05}. A tentative extrapolation of a TL model to SL 
region has also been done, and constraints have been found from the few available data \cite{Ba06}.

In TL region, data for EM FFs  exist over the NN threshold, up to 18 GeV$^2$ \cite{Am99} but a precise separation 
of the electric and magnetic contributions has not yet been possible, due to the low statistics. Moreover, FFs are complex 
quantities, and polarization experiments are necessary to determine their relative phase. Presently only the modulus of FFs 
is available, under the hypothesis $|G_E^N|^2=|G_M^N|^2$ or $G_E^N=0$. These data have been obtained in the 
reactions $e^++e^-\leftrightarrow \bar p+p$ (see Ref. \cite{Bai03} and references therein) and, more recently by the 
radiative return method \cite{Babar}, in the region over the kinematical threshold $s > 3.52$ GeV$^2$. The possible 
existence of an $N\bar N$ resonance under threshold has also been predicted as an explanation of the fact that TL FFs are 
larger than SL ones, at corresponding  $|q^2|$ values. In order to give quantitative predictions, for the cross section of the 
processes (\ref{eq:eqp}) and (\ref{eq:eqn}), it is necessary to know the value of FFs in the unphysical region. As data 
are not available, such estimation can be only done in the framework of models, based, for example, on VMD or on  dispersion 
relations, which predict several discontinuities due to meson resonances. Not all nucleon models give expressions for FFs 
which can be extended to time-like region, and not all nucleon models give a satisfactory description of all four nucleon FFs.

Following \cite{Etg05}, in the present calculation we use two models for electromagnetic FFs:  a model based on 
Ref. \cite{Ia73}, which firstly predicted the behavior of the proton electric FF as found from recent polarization 
experiments \cite{Jo00}, recently extended to TL region. This model, as all VMD inspired models, has poles in the 
unphysical region, in correspondence with the meson resonances which are taken into account. In order to have a smooth 
parametrization, we considered also a perturbative QCD (pQCD) inspired model prescription (corrected by dispersion 
relations \cite{Sol}) which reproduces the experimental data in TL region, does not show discontinuities and can be
 considered an 'average' expectation:
\begin{equation}
|G_E^N|=|G_M^N|=\displaystyle\frac{A(N)}
{q^4\left (\ln^2\displaystyle\frac{q^2}{\Lambda^2}+\pi^2\right )},
 ~q^2 > \Lambda ^2,
\label{eq:eqtp}
\end{equation}
where $\Lambda=0.3$ GeV is the QCD scale parameter and $A$ is fitted to the data.
This parametrization is taken to be the same for proton and neutron. The best fit is obtained with a parameter 
$A(p)$= 98 GeV$^4$ for the proton and $A(n)$= 134 GeV$^4$ for the neutron, which reflects the fact that in the TL region, 
neutron FFs are systematically larger than for the proton. In principle, this parametrization holds only for very large $q^2$ 
values, but, in practice, it reproduces the existing data quite well in the whole physical region. Evidently it is meaningless 
at small $q^2$, $(q^2<\Lambda^2)$, and it has not the good normalization properties for $q^2\to 0$.

Usually the data are shown in terms of the Sachs FFs, electric $G_E^N$, and magnetic $G_M^N$, which are related 
to the Pauli and Dirac FFs by the following relations:
$$G_E^N=F_1^N+F_2^N,~ G_M^N=F_1^N+\tau F_2^N,~\tau=q^2/(4M^2).$$.
They  correspond in a nonrelativistic limit (or in the Breit frame) to the Fourier transform of the charge density
(electric form factor $G_{E}$) and magnetization distribution (magnetic form factor $G_{M}$)
of the proton.

\begin{figure}
\begin{center}
\includegraphics[width=\textwidth]{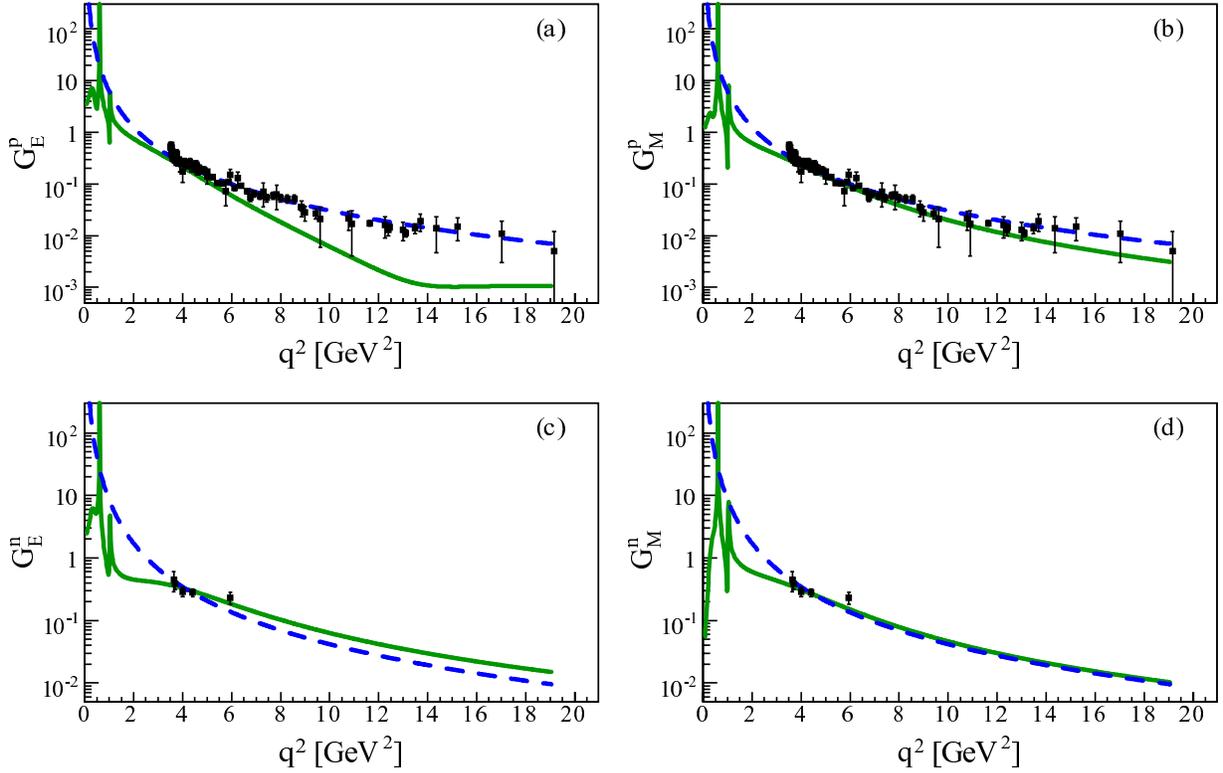}
\caption{\label{Fig:emff} (Color online) Nucleon electromagnetic FFs in time-like region: proton electric 
FF (a), proton magnetic FF (b), neutron electric FF (c), neutron magnetic FF (d). Data are from \protect\cite{Babar} 
and predictions from model \protect\cite{Ia73} (solid line), and from pQCD (dashed line).}
\end{center}
\end{figure}

The behavior of these FFs is shown in Fig. \ref{Fig:emff}, compared with the existing experimental data. 
For the neutron, the first and still unique measurement in TL region was done at Frascati, by the collaboration 
FENICE \cite{Ant98}. 

Concerning the pion FF, a reasonable description exists in the kinematical region of interest here, for a 
recent discussion see Ref. \cite{Br05}. For the sake of simplicity, we use here a $\rho$ meson saturated 
monopole-like parametrization, which takes a Breit Wigner form in TL region:
\be
F_{\pi}(q^2)=\displaystyle\frac{m_\rho^2}{m_\rho^2-q^2-i m_\rho \Gamma_\rho }.
\label{eq:frho}
\ee

Data on axial FFs in TL region do not exist, and they suffer in SL region from a model dependent derivation. 
In SL region, the nucleon axial FF, $G_A(q^2)$, for the transition $W^*+p\to n$ ($W^*$ is the virtual $W$-boson), 
can be described by the following simple formula \cite{Be02}:
\begin{equation}
G_A(q^2)=G_A(0)(1-q^2/m_A^2)^{-n}
\label{eq:eq6}
\end{equation}
with $m_A$ = 1.06 GeV, if $n=2$. A simple analytical continuation of this prescription to the TL region, presents 
a pole in the unphysical region. Therefore we used a 'mirror' parametrization from SL region: 
$$ FF^{(TL)}(|q^2|)= FF^{(SL)}(|q^2|).$$
Such a prescription is, in principle, valid only at very large $q^2$, since it obeys to asymptotic analytical properties of FFs \cite{Etg01}.

%%%%%%%%%%%%%%%%%%%%%%%%%%%%%%%%%%%%%%%%%%%%%%%%%%%%%%%%%%%%%%%%%%%%%%%%%%
\begin{figure}
\begin{center}
\includegraphics[width=12cm]{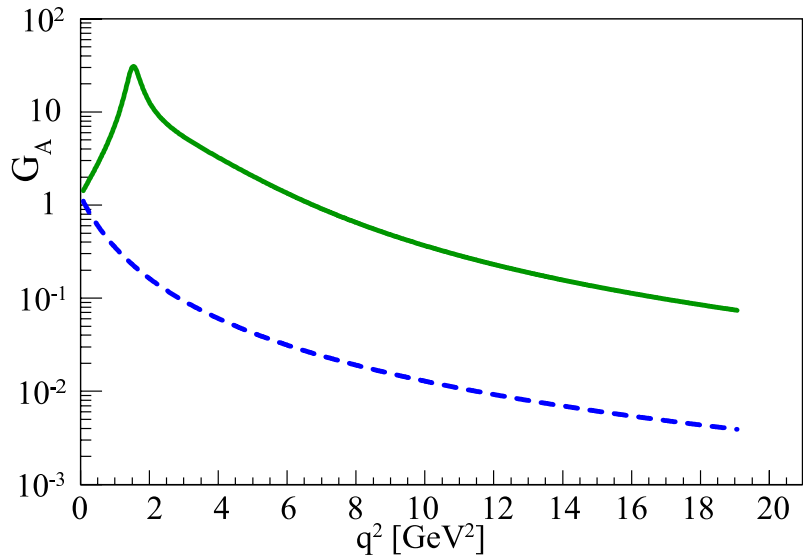}
\caption{\label{Fig:axff}  (Color online) Proton axial FF from VMD inspired model (solid line) and from a dipole extrapolation (dashed line).}
\end{center}
\end{figure}
%%%%%%%%%%%%%%%%%%%%%%%%%%%%%%%%%%%%%%%%%%%%%%%%%%%%%%%%%%%%%%%%%%%%%%%%%%%

For comparison, another parametrization is also used \cite{Etg07}, inspired from \cite{Ia73}:
\begin{equation}
G_A^{SL}(q^2)=d(q^2)G_A(0)\left [1-\alpha +\alpha \frac{m_A^2}{m_A^2-q^2} \right ],
\label{eq:axial}
\end{equation}

The parameter $\alpha =1.893\pm 0.02$ has been fitted on the available data, in the SL region, averaging the dispersion due to the
model dependent extraction of the data themselves, $m_A\simeq 1.235$ GeV is the mass of a light axial meson, and $d(q^2)=\left (1-\gamma q^2\right )^{-2}$
is the function describing the internal core of the nucleon. We take $\gamma$  as a fixed parameter, from the fit of
nucleon electromagnetic form factors :
$\gamma\simeq $0.25 GeV$^{-2}$. Let us note however that this value is not good from a $t$ channel point of view,
because it gives a pole in the physical region, $t_0=\frac{1}{\gamma}$=4 GeV$^{-2}$. 
In order to extend the expression (\ref{eq:axial}) to TL region, following \cite{Ia73},  a phase is added to the dipole term. 
Moreover the complex nature of the axial FF is insured by adding a width to the axial meson, 
and we replace the propagator as in  Eq. (\ref{eq:frho}).

The expression for the TL axial FF is therefore:
\begin{equation}
G_A(q^2)=d(q^2)G_A(0)\left [1-\alpha +\alpha \frac{m_A^2}{m_A^2-q^2-im_A\Gamma_A} \right ],
\label{eq:axial2}
\end{equation}
with $d(q^2)=\left (1-\gamma  e^{i\delta} q^2\right )^{-2}$, where $\Gamma_A$=0.140 GeV, and $\delta$=0.925.

The two models for $G_{A}$ used in the calculation of the cross section are shown in Fig.~\ref{Fig:axff} and differ 
by one order of magnitude. Moreover, en enhancement is expected from Eq. (\ref{eq:axial2}), in correspondance to the mass of the axial meson.

\section{Results}
The differential and integrated cross sections were calculated for several values of the antiproton energy and 
different choices of FFs described above.

\begin{figure}
\begin{center}
\includegraphics[width=0.49\textwidth]{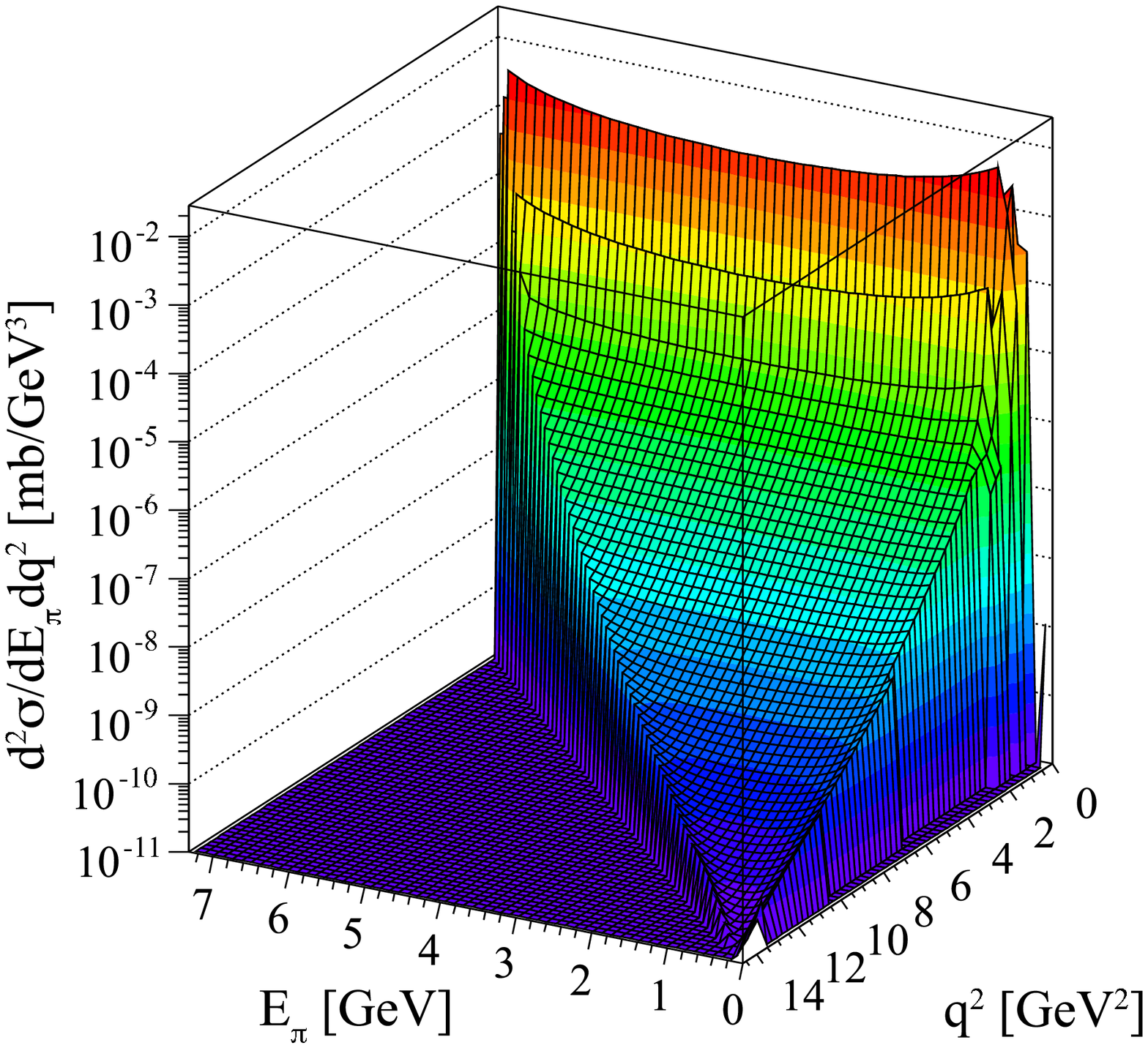}
\includegraphics[width=0.49\textwidth]{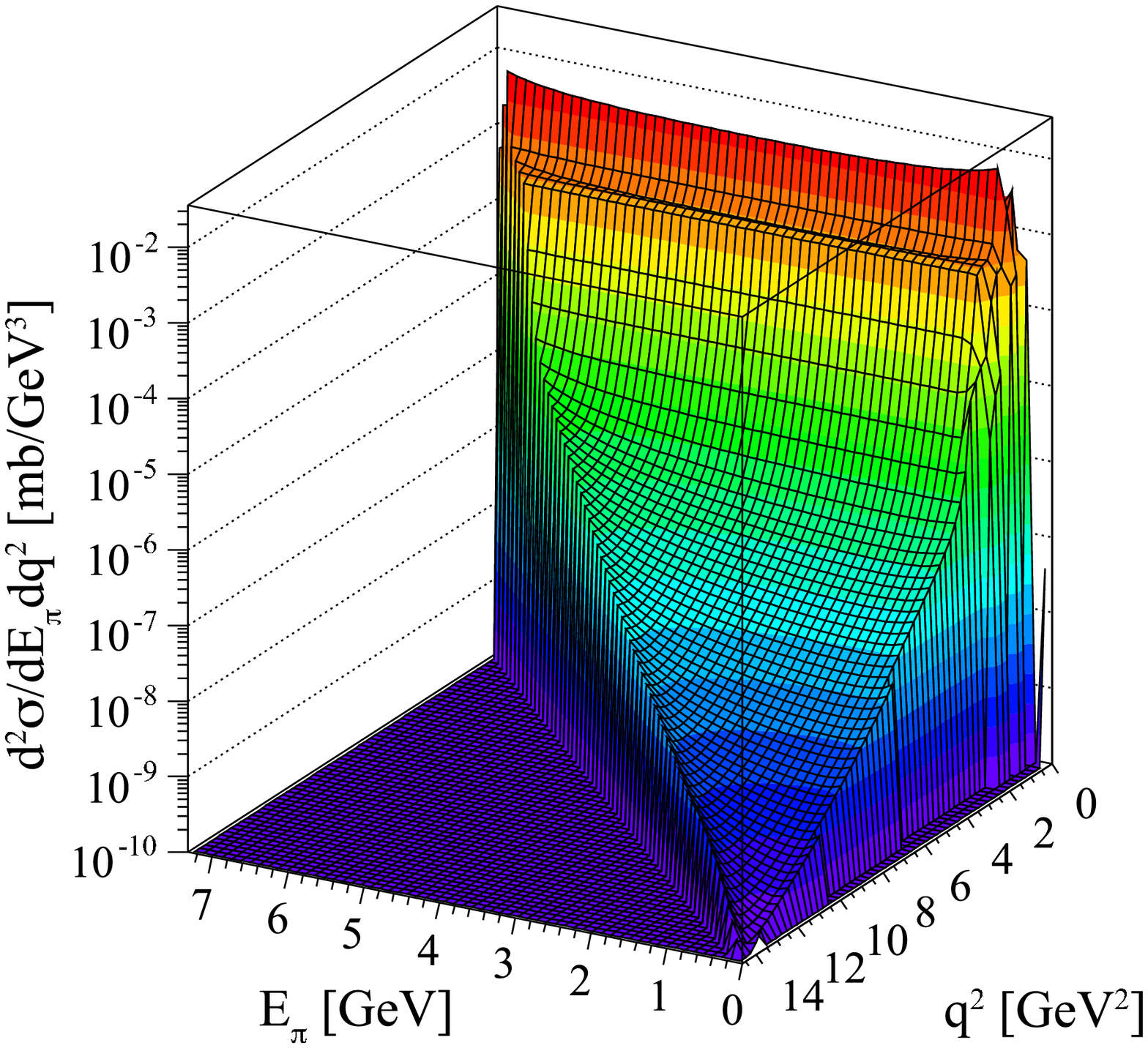}
\caption{\label{Fig:fig6}  (Color online) Left: Double differential cross section  for the reaction 
$\bar{p} +p \to \pi^0+\ell^++\ell^- $ as a function of $q^2$ and $E_{\pi}$, using FFs 
from \protect\cite{Ia73} for nucleon and (\protect\ref{eq:axial}) for axial FF. Right:
Same quantity as in the left plot for the reaction $ \bar{p} +n \to \pi^-+\ell^++\ell^- $. 
The kinematical constraints in the ($E_{\pi}, q^2$)-plane shown in Fig.~\ref{Fig:c2} are visible here.}
\end{center}
\end{figure}

\begin{figure}
\begin{center}
\includegraphics[width=0.49\textwidth]{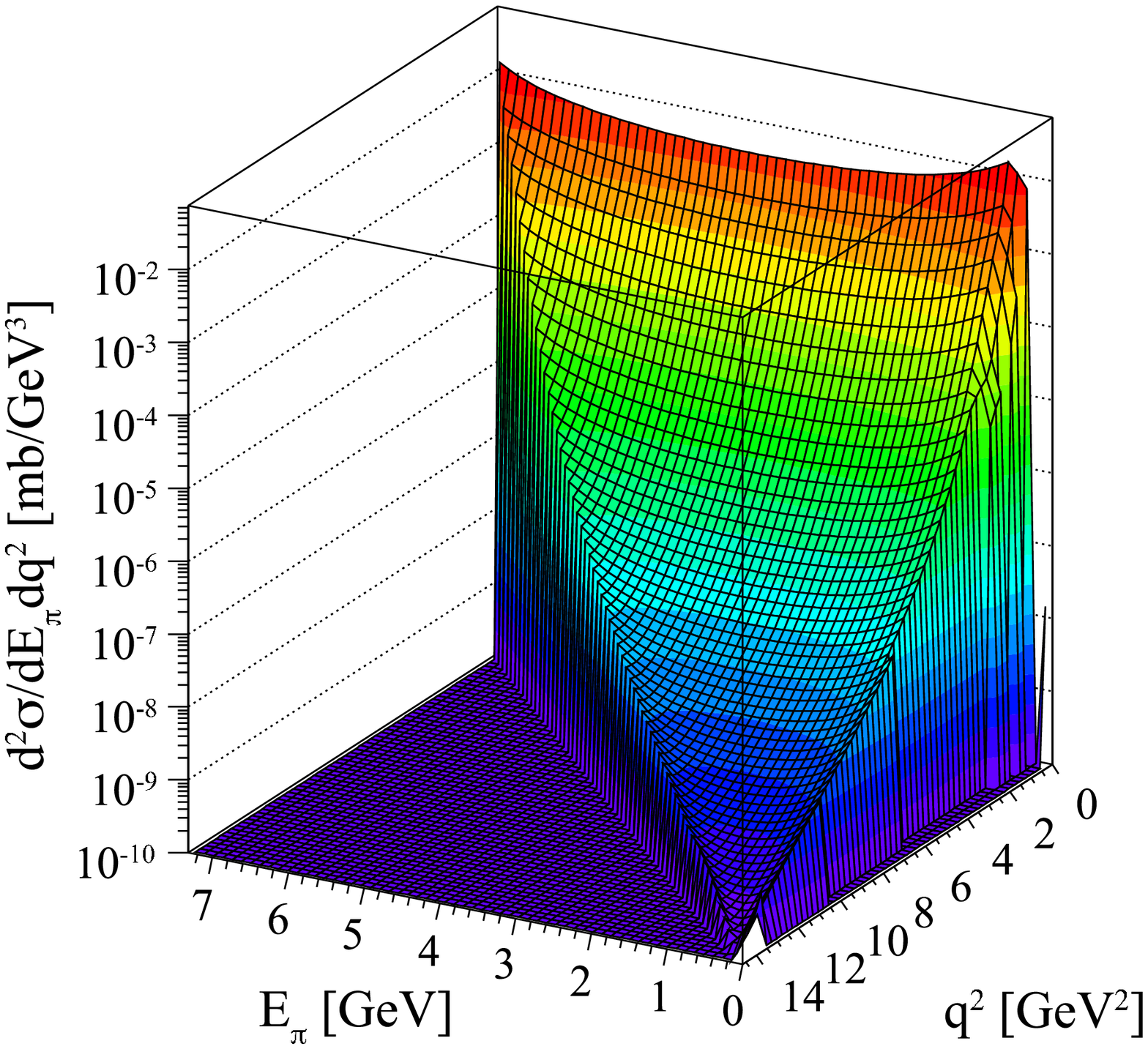}
\includegraphics[width=0.49\textwidth]{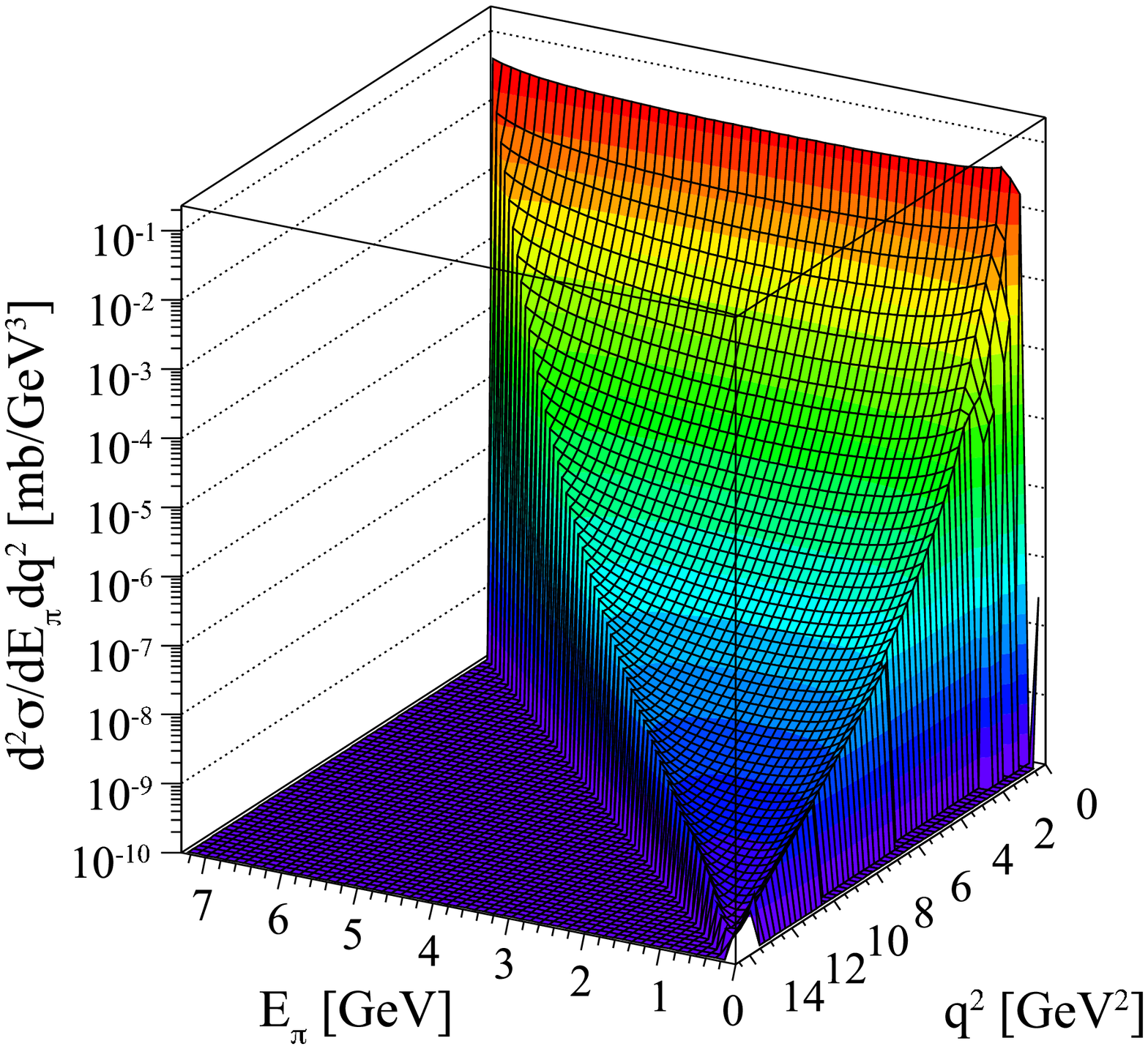}
\caption{\label{Fig:fig7}  (Color online) Left: Double differential cross section  for the reaction 
$\bar{p} +p \to \pi^0+\ell^++\ell^- $ as a function of $q^2$ and $E_{\pi}$,  
using pQCD inspired nucleon FFs and dipole axial FFs.
Right:
Same quantity as in the left plot for the reaction $ \bar{p} +n \to \pi^-+\ell^++\ell^- $.}
\end{center}
\end{figure}

The differential cross sections, Eq. (\ref{eq:dsig}), as a function of $E_{\pi}$ and $q^2$, Eq. (27a), are shown in 
Figs. \ref{Fig:fig6} and  \ref{Fig:fig7} for the reactions  
$ \bar{p} +p \to \pi^0+\ell^++\ell^- $ and $ \bar{p} +n \to \pi^-+\ell^++\ell^- $  and at $E$= 7 GeV$^2$. 
As one can see from the figures, the differential cross sections are large and measurable in a wide range of the considered 
variables. It is reasonable to assume that the region up to $q^2= 7$ GeV$^2$, at least, will be accessible by the experiments at FAIR.
%%%%%%%%%%%%%%%%%%%%%%%%%%%%%%%%%%%%%%%%%%%%%%%%%%%%%%%%%%%%%%%%%%%%%%%%%%%

%%%%%%%%%%%%%%%%%%%%%%%%%%%%%%%%%%%%%%%%%%%%%%%%%%%%%%%%%%%%%%%%%%%%%%%%%%%

%%%%%%%%%%%%%%%%%%%%%%%%%%%%%%%%%%%%%%%%%%%%%%%%%%%%%%%%%%%%%%%%%%%%%%%%%%%
%\begin{figure}
%\begin{center}
%\includegraphics[width=12cm]{figure08_dcs_Ep_q_pim_VMD.eps}
%\caption{\label{Fig:fig8}  (Color online) Same as Fig. \protect\ref{Fig:fig7} for the reaction $ \bar{p} +n \to \pi^-+\ell^++\ell^- $. }
%\end{center}
%\end{figure}
%%%%%%%%%%%%%%%%%%%%%%%%%%%%%%%%%%%%%%%%%%%%%%%%%%%%%%%%%%%%%%%%%%%%%%%%%%%%%

%%%%%%%%%%%%%%%%%%%%%%%%%%%%%%%%%%%%%%%%%%%%%%%%%%%%%%%%%%%%%%%%%%%%%%%%%%%
%\begin{figure}
%\begin{center}
%\includegraphics[width=12cm]{figure10_dcs_Ep_q_pim_pQCD.eps}
%\caption{\label{Fig:fig10}  (Color online) Same as Fig. \protect\ref{Fig:fig8} using pQCD inspired nucleon FFs and dipole axial FFs.}
%\end{center}
%\end{figure}
%%%%%%%%%%%%%%%%%%%%%%%%%%%%%%%%%%%%%%%%%%%%%%%%%%%%%%%%%%%%%%%%%%%%%%%%%%%

The discontinuities in the small $q^2$ regions are smoothed out by the steps chosen to histogram  the variables. 
However, depending on the resolution and the reconstruction efficiency, it will be experimentally possible to 
identify the meson and nucleon resonances.

%In order to illustrate the kinematical region in the plane $E_{\pi}$ - $\theta_{\pi}$, the double differential cross section 
%for the process  $ \bar{p} +n \to \pi^0+\ell^++\ell^- $ is shown in Fig. \ref{Fig:fig11}, for $s=7$ GeV$^2$.

%%%%%%%%%%%%%%%%%%%%%%%%%%%%%%%%%%%%%%%%%%%%%%%%%%%%%%%%%%%%%%%%%%%%%%%%%%%
%\begin{figure}
%\begin{center}
%\includegraphics[width=0.49\textwidth]{figure08_dcs_Ep_theta_pi0_VMD_gray.eps}
%\caption{\label{Fig:fig11} Double differential cross section for the process  $ \bar{p} +n \to \pi^0+\ell^++\ell^- $ 
%as a function $E_{\pi}$ and $\cos\theta_{\pi}$ using FFs from \protect\cite{Ia73} for nucleon and (\protect\ref{eq:axial}) 
%for axial FF (darker gray correspond to larger values).}
%\end{center}
%\end{figure}
%%%%%%%%%%%%%%%%%%%%%%%%%%%%%%%%%%%%%%%%%%%%%%%%%%%%%%%%%%%%%%%%%%%%%%%%%%%%%

\begin{figure}
\begin{center}
\includegraphics[width=0.49\textwidth]{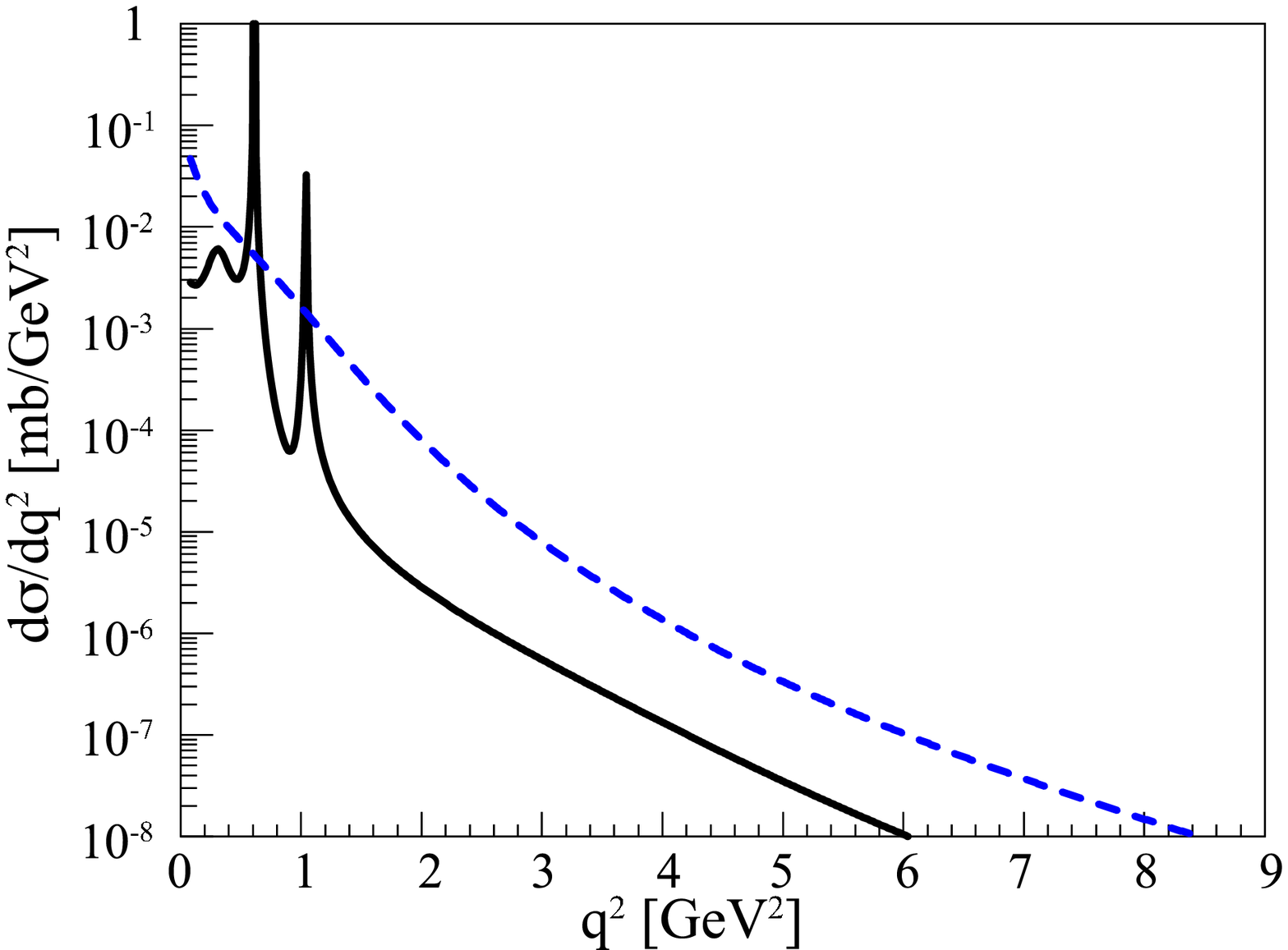}
\includegraphics[width=0.49\textwidth]{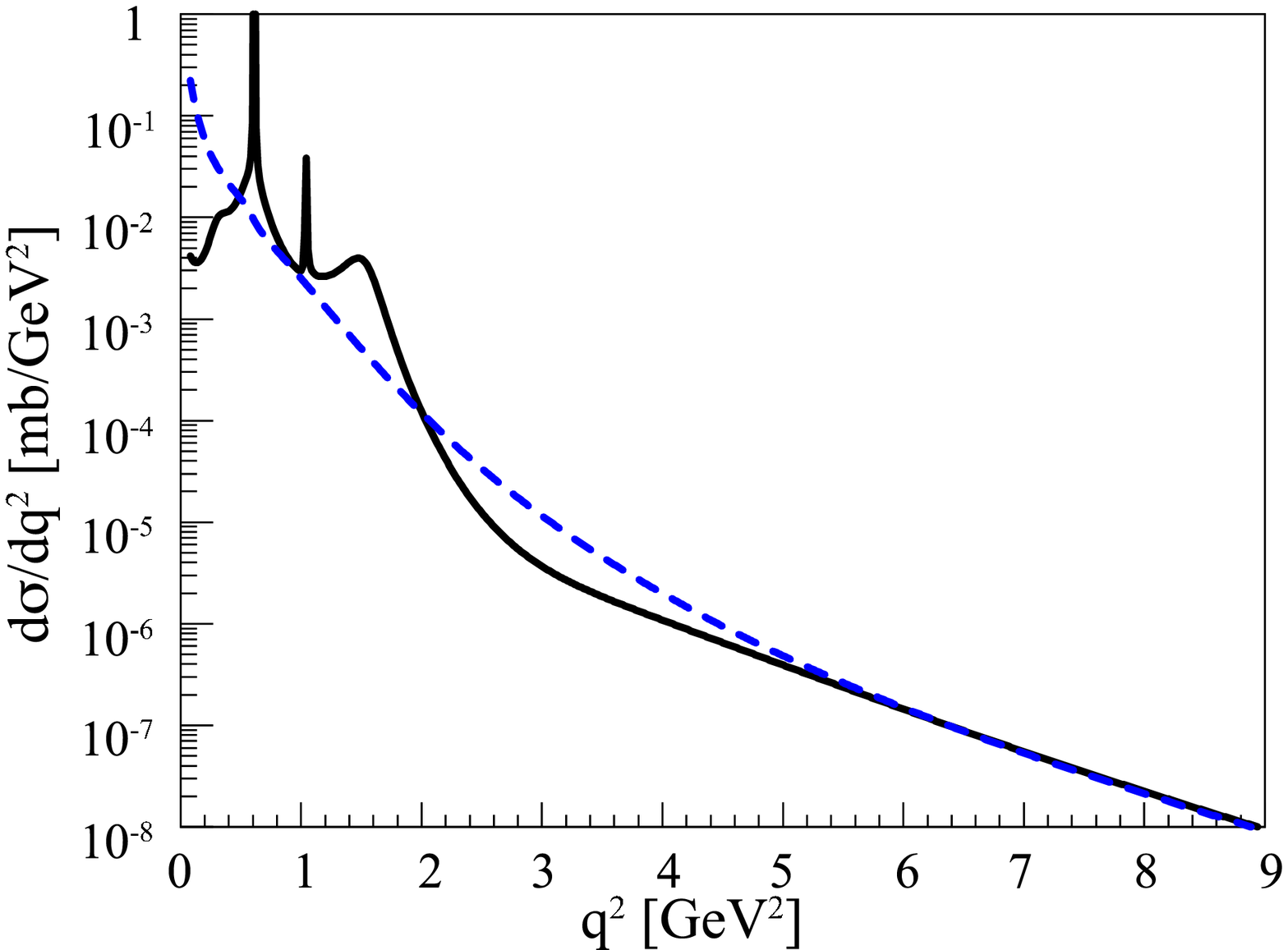}
\caption{\label{Fig:pps} (Color online) Left: Differential cross section for the process  $ \bar{p} +p \to \pi^0+\ell^++\ell^- $ 
as a function of $q^2$ , 
with FFs from \protect\cite{Ia73} for nucleon and (\protect\ref{eq:axial}) for axial FF (solid line) and with FFs 
from pQCD inspired nucleon FFs and dipole axial FFs (dashed line). Right: Same quantity as left,
for the reaction $ \bar{p} +n \to \pi^-+\ell^++\ell^- $. }
\end{center}
\end{figure}

The differential cross section as a function of $q^2$ can be obtained after integrating on the pion energy, 
Eq. (\ref{eq:dsig}), with the help of Eq. (27b):
\be
\frac{d\sigma^i}{dq^2}= 
\frac{\alpha^2}{6s\pi r}\frac{\beta(q^2+2\mu^2)}
{(q^2)^2}
\frac{M}{sr}
\int_{E_{\pi}^{min}}^{E_{\pi}^{max}}  {\cal D}^i dE_\pi,
\label{eq:eqint}
\ee
where the integration on the hadronic term is detailed in the Appendix.
The result of the calculation is shown in Fig.~\ref{Fig:pps}.  For charged pion 
production, the presence of the axial FF is the reason of a larger cross section as compared to the neutral pion case. 
For both reactions, again, the present calculation gives an integrated cross section of the order of several $\mu b$ in 
the unphysical region, for both choices of FFs.

\begin{figure}
\begin{center}
\includegraphics[width=0.49\textwidth]{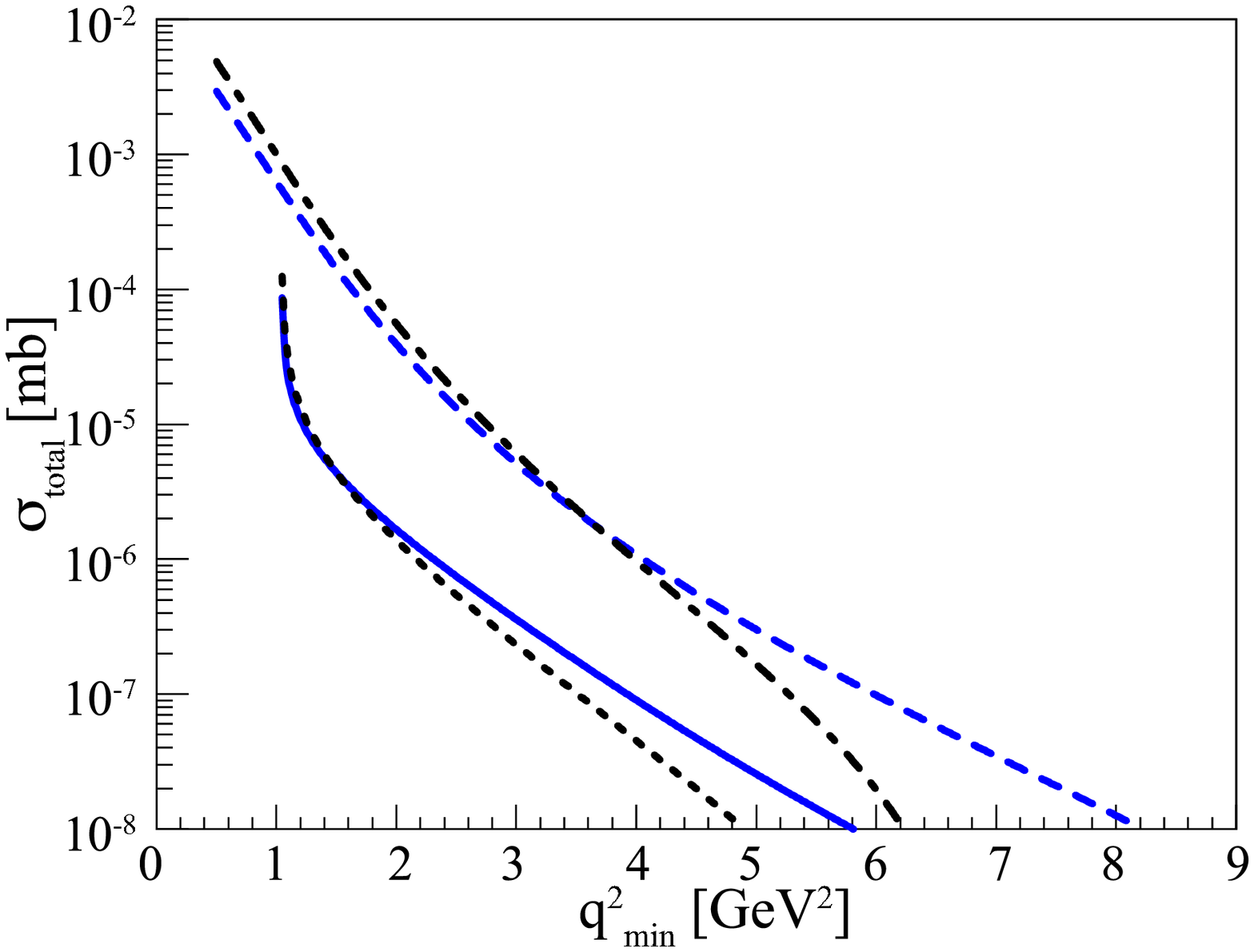}
\includegraphics[width=0.49\textwidth]{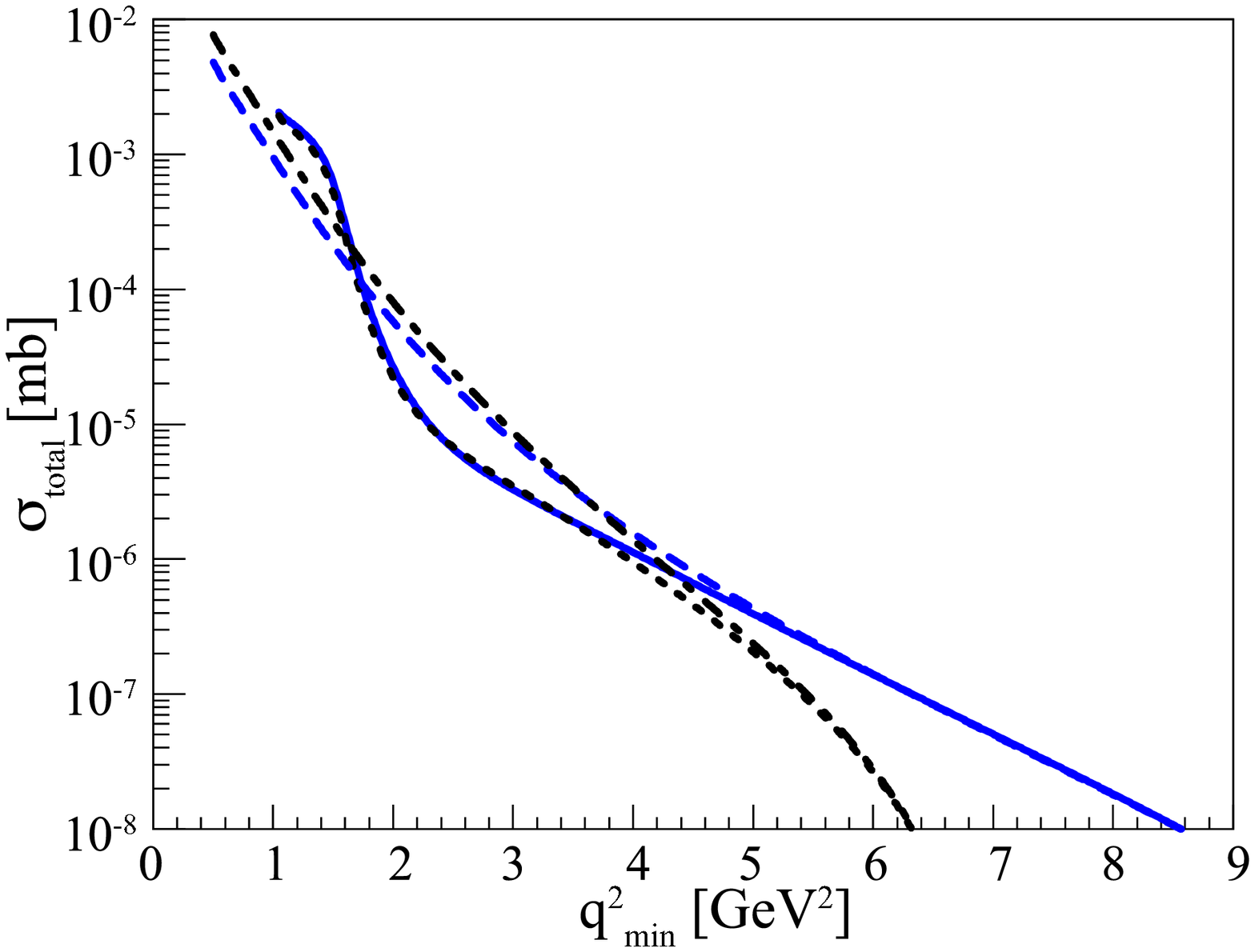}
\caption{\label{Fig:tcsp} (Color online) Left: Total cross section for the process  $ \bar{p} +p \to \pi^0+\ell^++\ell^- $
as a function of $q_{min}$ for different values of $s$ and different FFs: dotted (solid) line for $E$ =7(12) GeV$^2$,  
nucleon FFs from \protect\cite{Ia73} and axial FF from (\protect\ref{eq:axial}); dash-dotted (dashed) line for $E$ =7(12) GeV$^2$, 
pQCD inspired nucleon FFs and dipole axial FFs. Right: Same quantity as in the left plot,
for the process  $ \bar{p} +n \to \pi^-+\ell^++\ell^- $.  }
\end{center}
\end{figure}

The $q^2$ dependence is driven by the choice of FFs. In case of pQCD-like FFs, the behavior is smooth and similar 
for proton and neutron. In case of FFs from \cite{Ia73},  the resonant behavior due to $\rho$, $\omega$ and $\phi$ 
poles appears in the figures. The partial total cross section, integrated over $q^2$, Eq. (\ref{eq:eq23}), is shown in 
Fig.~\ref{Fig:tcsp}, as a function of $q_{min} \gg 4\mu^2$ for two different values of $s$, $s$=2 and 5 GeV$^2$. 
Evidently the calculations for the VMD FFs \cite{Ia73} have been done beyond the resonance region, 
due to the divergence of the integrals at the meson poles. 

%%%%%%%%%%%%%%%%%%%%%%%%%%%%%%%%%%%%%%%%%%%%%%%%%%%%%%%%%%%%%%%%%%%%%%%%%%%

%%%%%%%%%%%%%%%%%%%%%%%%%%%%%%%%%%%%%%%%%%%%%%%%%%%%%%%%%%%%%%%%%%%%%%%%%%%
%\begin{figure}
%\begin{center}
%\includegraphics[width=12cm]{figure13_integ_np.eps}
%\caption{\label{Fig:nps} (Color online) Same as Fig. \protect\ref{Fig:pps}, for the reaction $ \bar{p} +n \to \pi^-+\ell^++\ell^- $. }
%\end{center}
%\end{figure}
%%%%%%%%%%%%%%%%%%%%%%%%%%%%%%%%%%%%%%%%%%%%%%%%%%%%%%%%%%%%%%%%%%%%%%%%%%%
%%%%%%%%%%%%%%%%%%%%%%%%%%%%%%%%%%%%%%%%%%%%%%%%%%%%%%%%%%%%%%%%%%%%%%%%%%%
%\begin{figure}
%\begin{center}
%\includegraphics[width=12cm]{figure15_TCS_qmin_pn.eps}
%\caption{\label{Fig:tcsn} (Color online) Same as Fig. \protect\ref{Fig:tcsn} for the 
%process  $ \bar{p} +n \to \pi^-+\ell^++\ell^- $. }
%\end{center}
%\end{figure}

%%%%%%%%%%%%%%%%%%%%%%%%%%%%%%%%%%%%%%%%%%%%%%%%%%%%%%%%%%%%%%%%%%%%%%%%%%%
%\begin{figure}
%\begin{center}
%\includegraphics[width=12cm]{TCS_pp.eps}
%\caption{\label{Fig:tcsn} (Color online) Total cross section for the process  $ \bar{p} +p \to \pi^0+\ell^++\ell^- $ as a function of $s$ for different 
%values of $q^2_{min} =1.5 GeV^2$ (yellow line), $q^2_{min} =2.5 GeV^2$ (yellow line), and  $q^2_{min} =3.5 GeV^2$ (yellow line), f
%or FFs from \protect\cite{Ia73} for nucleon and (\protect\ref{eq:axial}) for axial FF (solid line) and with FFs from pQCD 
%inspired nucleon FFs and dipole axial FFs (dashed line).}
%\end{center}
%\end{figure}%%%%%%%%%%%%%%%%%%%%%%%%%%%%%%%%%%%%%%%%%%%%%%%%%%%%%%%%%%%%%%%%%%%%%%%%%%%
%\begin{figure}
%\begin{center}
%\includegraphics[width=12cm]{TCS_pn.eps}
%\caption{\label{Fig:tcsp} (Color online) Total cross section for the process  $ \bar{p} +n \to \pi^-+\ell^++\ell^- $ as a function of $s$ for different 
%values of $q^2_{min} =1.5 GeV^2$ (yellow line), $q^2_{min} =2.5 GeV^2$ (yellow line), and  $q^2_{min} =3.5 GeV^2$ (yellow line), 
%for FFs from \protect\cite{Ia73} for nucleon and (\protect\ref{eq:axial}) for axial FF (solid line) and with FFs from pQCD inspired 
%nucleon FFs and dipole axial FFs (dashed line). }
%\end{center}
%\end{figure}
Let us estimate the uncertainties inherent to our model assumptions. The off mass-shell effects were previously 
discussed for these particular reactions in \cite{Du95}. Large theoretical effort has been devoted to this problem 
in the past (Ref. \cite{Na87} and ref. therein), where it was shown that indeed off-mass shell effects can be large 
and have the effect to increase FFs. In the region of virtuality $p^2\le 0.5$ GeV $^2$ it was calculated that off mass 
shell effects modify FFs at the level of 3 \%. The $Q^2$-dependence of FFs is not changed significantly, when one 
of the particles goes off shell. It is interesting to note that the ratio of electric and magnetic nucleon FF is rather 
insensitive to off shell effects. Moreover ChPT arguments \cite{UM} support a smooth behavior of FFs as a 
function of the degree of nucleon virtuality $\delta\sim \left| \frac {p^2-M^2}{M^2}\right |^2 \le 2$, which 
does not exceed 10\%. For the present kinematics the virtuality involved varies in the interval,
 $2m_{\pi}/M<\delta<2-2m_{\pi}/M$. In the case of detection of soft pions in laboratory frame, 
 errors arising from off shell effects will decrease to $3\div 5$\%.

In the intermediate state, in principle, resonances can be excited. The estimation of contribution of the $\Delta$ resonance, 
compared to nucleon intermediate state, is of the order of $M_{\Delta}g_{\Delta N\pi}/g_{NN\pi}
\simeq 0.07$. Higher resonances have smaller contribution.

These considerations support an estimation on the precision of our model at the level of 10\%.

\section{Conclusion and perspectives}

The differential cross section for  the processes $\bar p + n\to \pi^- +\ell^-+\ell^+$ and $\bar p+ p\to \pi^0 +\ell^- + \ell^+$ 
has been calculated in the kinematical range which will be accessible in next future at FAIR. The main interest of these reactions is 
related to the possibility of measuring nucleon electromagnetic and axial FFs in the time-like and in the unphysical regions. 
As previously pointed out \cite{Re65,Du95}, varying the momentum of the emitted pion allows to scan the $q^2$ 
region of interest, keeping the beam energy fixed. 

In Ref. \cite{Du95} it was also noticed that in the lepton invariant mass squared distribution, a divergent 
term was present:
\be
\displaystyle\frac{d\sigma}{dq^2}\simeq \displaystyle\frac{[2F_1^v(q^2)+F_{\pi}(q^2)]^2}
{(q^2)^2},~q^2 \to 0,
\label{eq:eqre}
\ee
(in our notation the right hand side of Eq. (\ref{eq:eqre}) corresponds to ${\cal C}^2/(q^2)^2$). It was argued 
that the singularity at $q^2\to 0$ cancels due to $2F_1^v(0)=1$ and  $2F_{\pi}(0)=-1$. 
This compensation takes place if the following equality 
$g_{\pi\bar N N}(s)=g_{\pi\bar N N} (m_{\pi}^2)$ holds, which is verified for annihilation at rest. In the 
present work, one can not rely on such assumption and the validity of this relation has to be verified experimentally.

The detailed measurement of the double differential cross section, as a function of $q^2$ and $E_{\pi}$ allows 
in principle to extract all nucleon FFs which are involved in the considered reactions. A precise simulation
of different processes involving the production of a pion will be necessary with a study of the best kinematical
conditions in order to minimize background contribution. In particular the reaction 
$\bar{p} + p \to  \pi^0 + \pi^0$ has been identified as a potential source of background in the $e^+ e^-$-spectrum 
due to its Dalitz decay $\pi^0 \to e^+ e^- \gamma$.

The assumption about the validity of a generalized form of the  Golberger-Treiman relation, which allows to relate 
the pseudoscalar $g_{\pi NN}$ coupling constant to the axial nucleon FF can be experimentally verified in case of 
small invariant mass of the lepton pair.
The possibility of measurement of heavy negatively charged pions ( $\pi'$ -radial excitation of pion, $M'=1300$ MeV) 
in the process of $\pi^-$ production can be taken into account in the following way.
The matrix element for the excitation of a resonant state for the  virtual charged pion in the intermediate state, $M_{res}$, 
can be written as follows:
\be
M_{res}=\displaystyle\frac{4\pi\alpha}{q^2}
\displaystyle \frac{ g_{\pi N\bar N}}{s-M'^2+iM'\Gamma '}
\bar{v}(p_1)\gamma_5 u(p_2)\lambda(q^2) \left (q_{\mu}-
\displaystyle\frac{q^2}{qq_{\pi} }q_{\pi\mu} \right )J_{\mu}
\ee
The corresponding cross section is:
\be
d\sigma_{res}= \frac{\alpha^2}{12 r\pi} 
\frac{|\lambda(q^2)g_{\pi N\bar N}|^2}{(s-M'^2)^2+M'^2\Gamma'^2}
\frac{d^3q_{\pi}}{2\pi E_{\pi}}.
\label{eq:eqs}
\ee
where $\lambda(q^2)$ is the transition from factors for the vertex $\pi ' \pi\gamma^*$.
We do not consider processes involving vector mesons, $\Delta$ resonances and higher excited nucleon states, 
estimating that their contribution does not exceed 10\%.

In case of multi-pion production, the quantity $s_1=(p_1+p_2-q)^2-m_{\pi}^2$ becomes positive. Varying $s_1$ at fixed 
beam energy, by changing $q^2$ and $\theta_{\pi}$, it is in principle possible to identify and study other mechanisms, 
as the excitation of heavy pion resonances, $\pi'$, or the possible presence of a $N\bar N$ 'quasi-deuteron' state 
under the kinematical threshold for $p\bar p$ annihilation in two leptons. The study of multipion production 
will be the subject of a forthcoming publication.

%%%%%%%%%%%%%%%%%%%%%%%%%%%%%
\section{Acknowledgments}
%%%%%%%%%%%%%%%%%%%%%%%%%%%%%
The authors are thankful to L. Lipatov for critical remarks on the
manuscript and interesting discussions. Useful discussion with S.~Scherer,  H.~W.~Hammer and G. I. Gakh are acknowledged. Two of us (E. A. K. and C. A.) are grateful to DAPNIA/SPhN, Saclay, where this work was done. The Slovak Grant Agency for Sciences VEGA is acknowledged by C.A. for support under Grant N. 2/4099/26.
%%%%%%%%%%%%%%%%%%%%%%%%%%%%%
\section{Appendix}
%%%%%%%%%%%%%%%%%%%%%%%%%%%%%

In this appendix we give the explicit expression of the coefficients entering in the 
calculation of the cross section, as well as useful integrals.

Let us define $q^2$-dependent terms, which contain FFs and the necessary constants:

$$f_a(s)=F_{\pi}(q^2)G_{\pi N\bar N}(s), ~
f_{iN}(q^2)=g(m_{\pi}^2)F_i^N(q^2),~i=1,2, ~N=n,p,$$
$${\cal C}(s)=f_a(s) - f_{1p}(q^2) + f_{1n}(q^2),$$
and the quantity:
\be
X=\displaystyle\frac{p_1 q_{\pi}}{p_2 q_{\pi}}=(s-q^2)/(2 ME_{\pi})-1.
\label{eq:eqx}
\ee

Let us write the expressions for the hadronic part of the matrix element (\ref{eq:eqcd}):

{\bf - for the process $p +\bar p \rightarrow \ell^+ + \ell^-+\pi^0$}
\begin{equation}
\mathcal{D}^0=|f_{2p}|^2 \left[ \frac{E-M}{M} - \frac{1}{2}
\left( 1-\frac{q^2}{4M^2} \right) \frac{(1-X)^2}{X} \right]+|f_{1p}-f_{2p}|^2\frac{(X+1)^2}{X},
\end{equation}

{\bf - for the process $n +\bar p \rightarrow \ell^+ + \ell^- +\pi^-$}
\be
\mathcal{D}^-=\frac{1}{4}\left[\sum_i C_{i,i} |f_i|^2 +  2\sum_{j,k  ; j< k} C_{j,k} Re (f_j f_k^*) + 
\frac{2|{\cal C}|^2s}{q^2}\right],é i,j,k=1p,2p,1n,2n,a.
\ee
The explicit form of the coefficients is:
\begin{eqnarray}
C_{1p,1p}&=&4 X,~
C_{2p,2p}= \frac{s}{M^2} \left(1+\frac{q^2}{2s}X\right),~
C_{1p,2p}=-3 \left(1 +X\right),~
C_{a,a}=\frac{2q^2}{s}-4,
\nonumber\\
C_{1n,1n}&=&4 \frac{1}{X} ,~
C_{2n,2n}=\frac{s}{M^2}\left ( 1 + \frac{q^2}{2sX} \right ),~
C_{1n,2n}=-3 \left( 1 +\frac{1}{X}   \right),~
\nonumber \\
C_{a,1p}&=&2,~
C_{a,2p}= \left( 1-\frac{q^2}{s} \right)\left(1+ X\right),~
C_{1p,1n}=4,~
C_{1p,2n}= \left( \frac{1}{X} -2X- 1 \right ),
\nonumber \\
C_{2p,2n}&=& \left( 2 +\frac{2}{X} -\frac{q^2}{2M^2}+X \right),~
C_{2p,1n}= \left (X -\frac{2}{X} -1 \right),~
\nonumber \\
C_{a,1n}&=&-2,~
C_{a,2n}=- \left( 1-\frac{q^2}{s} \right)\left(1+ \frac{1}{X}\right).
\end{eqnarray}
The structure of ${\cal D}_i$ allows to select the terms which depend on the pion energy. 
It is straightforward to perform an analytical integration on the pion energy, using the following integrals:
\be
\int_{E_{\pi}^{min}}^{E_{\pi}^{max}}  \displaystyle\frac{dE_{\pi}}{M}
=\displaystyle\frac{r(s-q^2)}{2M^2}=rb;
~b= \displaystyle\frac{s-q^2}{2M^2},
\ee
\be
\int_{E_{\pi}^{min}}^{E_{\pi}^{max}}  \displaystyle\frac{dE_{\pi}}{M} X=
\int_{E_{\pi}^{min}}^{E_{\pi}^{max}}  \displaystyle\frac{dE_{\pi}}{M}
\displaystyle\frac{1}{ X}= \displaystyle\frac{s-q^2}{2 M^2}\left [ \ln 
\displaystyle\frac{1+r}{1-r} - r\right ]=b(\ell -r);
\ee
with $r$ given in (\ref{eq:eqr}) and  $\ell= \ln[(1+r)/(1-r)]$. The result of the 
integration on the pion energy is:

{\bf - for the process $p +\bar p \rightarrow \ell^+ + \ell^-+\pi^0$}
\be
\int_{E_{\pi}^{min}}^{E_{\pi}^{max}} \mathcal{D}^0 ~\frac{ dE_{\pi}}{M}={b}
\left \{ 2|f_{1p}-f_{2p}|^2\ell 
+|f_{2p}|^2 \left[ \frac {E-M}{M} r +\left(1-\frac{q^2}{4M^2} \right) \left(2r -\ell \right) \right ] \right \}
\ee

{\bf - for the process $p +\bar p \rightarrow \ell^+ + \ell^-+\pi^0$}
\be
\int_{E_{\pi}^{min}}^{E_{\pi}^{max}} 
\mathcal{D}^-~\frac{ dE_{\pi}}{M}=\frac{b}{4}\left[\sum_i K_{i,i} |f_i|^2 + 2\sum_{j,k ; j< k} K_{j,k} Re (f_j f_k^*) + |{\cal C}|^2\frac{2rs}{q^2} \right],
\ee
where 
\ba
K_{1p,1p}&=&4 \left( \ell -r \right),~
K_{2p,2p}= \frac{s}{M^2} \left(r+\frac{q^2}{2s}\left( \ell-r \right)\right),~
\nonumber\\
K_{1p,2p}&=&-3 \ell,~
K_{a,a}=\left(\frac{2q^2}{s}-4\right) r,~
\nonumber\\
K_{1n,1n}&=&4  \left( \ell -r \right),~
K_{2n,2n}= \frac{s}{M^2} \left( r + \frac{q^2}{2s}\left( \ell -r \right) \right),~
K_{1n,2n}=-3 \ell,~
\nonumber\\
K_{a,1p}&=&2 r,~
K_{a,2p}=\left( 1-\frac{q^2}{s} \right)\ell,~
K_{1p,1n}=4 r,~
K_{1p,2n}=- \ell,~
\nonumber\\
K_{2p,2n}&=& \left [  3\ell - \left ( 1+\frac{q^2}{2M^2} \right ) r  \right ],~
K_{2p,1n}=- \ell,~
\nonumber\\
K_{a,1n}&=&-2 r,~
K_{a,2n}=- \left( 1-\frac{q^2}{s} \right)\ell,~
\ea

\end{document}